\documentclass[11pt]{article}
\usepackage[margin=1in]{geometry}
\usepackage{booktabs,array,amsmath,amssymb,cite,url,xurl,graphicx,xcolor,tabularx,float,placeins,microtype}
\usepackage{siunitx}
\usepackage[hidelinks]{hyperref}
\usepackage[caption=false,font=small]{subfig}
\usepackage{authblk}
\title{Exascale Multi-Task Graph Foundation Models for Imbalanced, Multi-Fidelity Atomistic Data}
\author[1]{Massimiliano Lupo Pasini\thanks{Corresponding author: lupopasinim@ornl.gov}}
\author[2]{Jong Youl Choi}
\author[2]{Kshitij Mehta}
\author[3]{Richard Messerly}
\author[4]{Rylie Weaver}
\author[4]{Linda Ungerboeck}
\author[3]{Isaac Lyngaas}
\author[1]{Benjamin Stump}
\author[5]{Ashwin M. Aji}
\author[5]{Karl W. Schulz}
\author[5]{Jord\`a Polo}
\affil[1]{Computational Sciences and Engineering Division, Oak Ridge National Laboratory, Oak Ridge, Tennessee, USA}
\affil[2]{Computer Science and Mathematics Division, Oak Ridge National Laboratory, Oak Ridge, Tennessee, USA}
\affil[3]{National Center for Computational Sciences Division, Oak Ridge National Laboratory, Oak Ridge, Tennessee, USA}
\affil[4]{Bredesen Center for Interdisciplinary Research and Graduate Education, University of Tennessee, Knoxville, Tennessee, USA}
\affil[5]{Advanced Micro Devices, Inc., USA}
\date{}

\begin{document}
\maketitle

\begin{abstract}
{\textcolor{black}{We present an exascale workflow for atomistic graph foundation models using HydraGNN. Multi-task learning enables joint training across 16 heterogeneous first-principles datasets representing over 467 million unique structures (over 544 million cumulative branch records) and covering 85+ elements. A shared graph neural network backbone with dataset-specific prediction heads accommodates the strongly imbalanced, multi-source, multi-fidelity corpus, supported by scalable ADIOS2/DDStore data management. Composition-conditioned weighting combines source-specific predictions into a single deployable model. Six large-scale hyperparameter optimization campaigns, followed by resource-aware continual training on 2,048 Frontier nodes, identify PaiNN as the lead architecture under a fixed computational budget. We evaluate scalability across Frontier, Aurora, and Perlmutter and characterize transfer across twelve downstream tasks, including comparisons with pretrained UMA and MACE models under different adaptation strategies. These results reveal task-dependent transfer benefits and highlight the importance of matching adaptation strategies to downstream applications. Finally, we assess inference optimizations and BF16/FP32/FP64 precision tradeoffs, achieving the screening of 1.1 billion atomistic structures in approximately 50 seconds on Frontier. Together, these results establish a scalable framework for multi-source atomistic foundation model training, evaluation, and high-throughput deployment, thereby allowing for a fast and reliable exploration of vast chemical design spaces that are otherwise inaccessible to first-principles methods.
}}

\end{abstract}


\begin{quote}\footnotesize
This manuscript has been authored in part by UT-Battelle, LLC, under contract DE-AC05-00OR22725 with the U.S. Department of Energy (DOE). The U.S. government retains and the publisher, by accepting the article for publication, acknowledges that the U.S. government retains a nonexclusive, paid-up, irrevocable, worldwide license to publish or reproduce the published form of this manuscript, or allow others to do so, for U.S. government purposes. DOE will provide public access to these results of federally sponsored research in accordance with the \href{http://energy.gov/downloads/doe-public-access-plan}{DOE Public Access Plan}.
\end{quote}

\section{Introduction}
Accurate atomistic simulation is a core tool for materials discovery, chemistry, and molecular design. Density functional theory (DFT) and related first-principles methods provide high-quality labels for energies, forces, and other observables, but their computational cost limits the number of candidate structures that can be evaluated directly. Machine-learning interatomic potentials (MLIPs) and atomistic graph foundation models (GFMs) address this bottleneck by learning transferable representations from large electronic-structure datasets and amortizing the cost of training over subsequent simulation, screening, and fine-tuning tasks \cite{chen2022m3gnet,deng2023chgnet,batatia2024foundation,zhang2024dpa2largeatomicmodel,Takamoto2022}.

The still unaddressed challenge of scientific machine learning (ML) is that no single dataset or electronic-structure setting is uniformly reliable across organic, inorganic, and hybrid systems. Therefore, building truly general GFMs requires jointly training on multi-source, multi-fidelity datasets generated with different approximation theories and parameterizations (e.g., different functionals and basis sets), which introduces inconsistent energy references, label-noise heterogeneity, and optimization imbalance across tasks \cite{shiota2024tamingmultidomainfidelitydata,zhang2024dpa2largeatomicmodel,Jacobson2023,Allen2024,messerly_multi-fidelity_2025,chen_one_2025}. The 16 datasets used in this work are detailed in Section~\ref{sec:dataset_integration} and Table~\ref{tab:datasets_overview}. 

A further challenge is that energy-conserving MLIPs rely on force labels computed as energy gradients with respect to atomic positions; the resulting derivative-based learning signals amplify numerical errors and interact with reduced precision and distributed reductions. We examine machine-precision effects at inference and fine-tuning using the lead checkpoint selected from large-scale continual training. This clarifies how machine precision affects deployment-time and accuracy in scientific ML workflows that embed physical structure through differential operators \cite{LupoPasini2025HydraGNN}.

In this work, we address these challenges at leadership-computing scale by training a multi-task learning (MTL) graph neural network (GNN) architecture implemented in HydraGNN \cite{hydragnn4,LupoPasini2025HydraGNN} over \textcolor{black}{16 open, multi-source and multi-fidelity dataset branches representing over 467 million unique structures and over 544 million cumulative branch records} and spanning organic, inorganic, catalytic, hybrid, and polymer systems. Dataset-specific prediction heads are used to separate heterogeneous supervision while preserving a common message-passing backbone. The training workflow integrates ADIOS2 and DDStore for scalable data movement, supports sharded and task-parallel execution, and uses DeepHyper to search six equivariant message-passing families under a common production-oriented protocol.

The paper makes five contributions. First, it shows large-scale joint training over a chemically and numerically heterogeneous open corpus rather than a single dominant source. Second, it couples multi-task learning with a composition-conditioned branch-weighting model that converts source-specific output heads into a deployable predictor for previously unseen compositions. Third, it treats architecture selection as a joint accuracy--cost problem by combining large-scale hyperparameter optimization with fixed-budget continual training. Fourth, it establishes workflow portability across Frontier, Aurora, and Perlmutter and benchmarks downstream transfer against representative pretrained atomistic models---UMA, MACE-OFF23, MACE-MH-1, and MACE-MP-0---across zero-shot, frozen, low-rank, and full-adaptation regimes. This comparison identifies how the preferred model and adaptation strategy change with chemistry, dataset size, target, and pretraining/downstream alignment. Fifth, it develops and measures a sequence of increasingly refined inference optimizations, together with BF16/FP32/FP64 sensitivity studies, culminating in billion-structure screening at near-full Frontier scale.

\section{Related Work}
Recent atomistic ML has advanced along two complementary directions: increasingly expressive equivariant architectures and increasingly broad pretrained potentials. Equivariant and geometry-aware models such as GemNet, MACE, Allegro, Equiformer, and EquiformerV2 improve representation of local atomic environments and have produced strong results on large catalyst and materials benchmarks \cite{gasteiger2021gemnet,gasteiger2022gemnetoc,batatia2022mace,musaelian2022allegro,liao2022equiformer,liao2023equiformerv2}. Universal or broadly pretrained MLIPs, including M3GNet, CHGNet, MACE-OFF23, MACE-MP-0, MACE-MH-1, UMA, DPA-2, and PFP, extend this direction toward transfer across molecular and materials classes and downstream tasks \cite{chen2022m3gnet,deng2023chgnet,kovacs2025maceoff,batatia2024foundation,batatia_cross_2025,wood2025uma,zhang2024dpa2largeatomicmodel,Takamoto2022}.

UMA's expert-routing strategy is complementary to our core MTL design goal: UMA-style mixtures of experts mainly specialize across chemistry/composition regions \cite{wood2025uma}, whereas our approach specializes prediction heads across datasets and fidelity levels while retaining a shared message-passing foundation \cite{shiota2024tamingmultidomainfidelitydata,hydragnn4}. 

While MACE-MH-1 employs an MTL formulation similar to ours, its training pipeline proceeds in the reverse order \cite{batatia_cross_2025}. The underlying foundation MLIP, MACE-OMAT-1, is first pre-trained on the single-source, single-fidelity OMat24 dataset and only afterward fine-tuned on multi-source, multi-fidelity datasets such as OMol25, with replay of the original training data to limit catastrophic forgetting. In contrast, our workflow exposes the model to heterogeneous multi-source data from the start, thereby avoiding the sequential-update setting in which catastrophic forgetting can arise. 

Most published results optimize primarily for predictive error on fixed benchmarks \cite{riebesell2024matbench}, while reporting limited joint analysis of accuracy, time-to-solution, and computational cost-to-solution under identical training protocols. Reported training corpora often emphasize either single-source organic or inorganic domains, or rely on small or moderate multi-source training data, leaving unresolved how to robustly co-train on large volumes of imbalanced, multi-source, multi-fidelity data describing organic, inorganic, and hybrid systems at leadership-system scale \cite{takeda2023multimodal,beaini2024towards,barrosoluque2024openmaterials2024omat24,shoghi2024from}. \textcolor{black}{Our earlier large-scale study characterized the challenges of aggregating heterogeneous, strongly imbalanced atomistic datasets \cite{LupoPasini2025HydraGNN}. A subsequent controlled five-dataset study compared independent single-source models, merged single-stream training with one shared output decoder, and multi-task learning with dataset-specific heads; it reported lower energy-prediction errors for the multi-task configuration than for single-stream training across all five evaluated sources \cite{LupoPasini2026MultiTaskParallelism}. These results motivate the multi-task design used in the present exascale study.} Direct comparisons of predictive accuracy and computational performance across existing atomistic foundation models remain challenging because published studies use different training datasets, numerical precision settings, computing architectures, and performance measurement protocols.

\section{Methods and System Design}
\subsection{Multi-Source, Multi-Fidelity Training Corpus}\label{sec:dataset_integration}
\begin{figure}[H]
\centering
\includegraphics[width=0.96\textwidth]{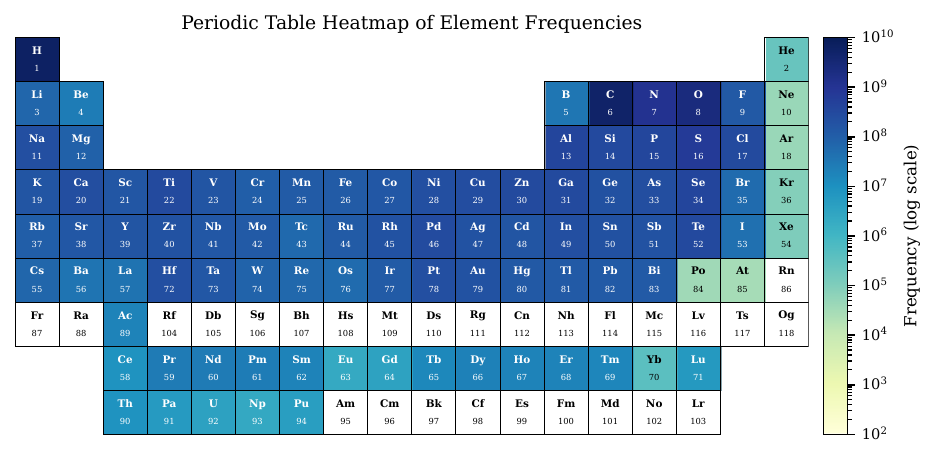}
\caption{Elemental coverage of the 16 open-source datasets used for pretraining. Each periodic-table cell is labeled by element symbol and atomic number, while color intensity represents the aggregate element frequency across the combined corpus on a logarithmic scale. The visualization highlights both the broad chemical coverage of the training data and the strong imbalance in elemental occurrence.}
\label{fig:periodic_table_coverage}
\end{figure}

The corpus contains \textcolor{black}{16 dataset branches totaling over 544 million branch records corresponding to over 467 million unique atomistic structures, covering 85+ elements (Table~\ref{tab:datasets_overview})}, including organic molecules, inorganic crystals, catalytic interfaces, hybrid compounds, and polymers. The data were produced by independent first-principles pipelines and therefore differ in exchange--correlation approximations, chemical coverage, label statistics, and dataset size. Figure~\ref{fig:periodic_table_coverage} visualizes the aggregate elemental coverage of the complete pretraining corpus. The logarithmic color scale makes both the breadth of the chemical space and the strong imbalance in elemental occurrence visible: the combined datasets cover a large fraction of the periodic table, while frequently occurring light elements and many transition metals appear orders of magnitude more often than sparsely represented elements. This imbalance motivates the multi-task and sampling strategies used in the training workflow. \textcolor{black}{The OMol25 neutral-singlet and non-neutral/non-singlet subsets are included in the full OMol25 dataset but are trained as separate supervision branches. Their structures therefore contribute to both the full-dataset branch and the corresponding subset branch. Table~\ref{tab:datasets_overview} distinguishes these repeated branch records from unique structures.} \textcolor{black}{For the overlapping OMol25 branches, data partitions were kept consistent across the complete dataset and its two subsets: a structure assigned to validation or testing in one branch was not used for training in another. Thus, the intentional reuse of records across branches did not introduce cross-partition leakage between training and evaluation.}

\begin{table}[t]
\caption{Multi-source, multi-fidelity first-principles datasets used in this work.\label{tab:datasets_overview}}
\centering
\small
\renewcommand{\arraystretch}{1.12}
\setlength{\tabcolsep}{3pt}
\begin{tabularx}{\textwidth}{@{}>{\raggedright\arraybackslash}p{0.30\textwidth} >{\raggedright\arraybackslash}p{0.22\textwidth} >{\raggedright\arraybackslash}X r@{}}
\toprule
\textbf{Dataset} & \textbf{Compounds} & \textbf{Approximation} & \textbf{Samples} \\
\midrule
QM7-X \cite{qm7X}                          & Organic molecules                  & DFT (PBE0+MBD)                 & 8,390,474   \\
QCML-DFT \cite{qcml2025}                  & Organic molecules                  & DFT (PBE0)                     & 33,496,167  \\
Transition1x \cite{schreiner2022transition1x} & Organic molecules               & DFT ($\omega$b97x/6-31G(d))   & 9,644,740   \\
ANI-1x \cite{Smith2020ANI1ccx}            & Organic molecules                  & DFT ($\omega$b97x/6-31G(d))   & 4,956,005   \\
$\nabla^2$-DFT \cite{khrabrov2024nabla2dft} & Organic molecules                & DFT ($\omega$b97x-D/def2-SVP) & 10,392,954  \\
Materials Project Trajectories (MPTrj) \cite{mptrj} & 3D bulk materials            & DFT (PBE/PBE+U)                & 1,580,395   \\
Alexandria Materials Database \cite{alexandriapbepaper} & 3D inorganic compounds   & DFT (PBE)                      & 5,364,523   \\
Open Catalyst 2020 (OC20) \cite{chanussot2021openCatalyst20} & Catalyst slabs        & DFT (RPBE/PAW)                 & 134,933,884 \\
Open Catalyst 2022 (OC22) \cite{tran2023openCatalyst22} & Catalyst slabs            & DFT (PBE(+U)/PAW)              & 8,366,325   \\
Open Catalyst 2025 (OC25) \cite{sahoo2025oc25} & Catalyst slabs                   & DFT (PBE(+U)/PAW)              & 7,599,142   \\
Open Direct Air Capture 2023 (ODAC23) \cite{sriram2024odac23} & Hybrid compounds      & DFT (PBE-D3(BJ)/PAW)           & 36,710,860  \\
Open Materials 2024 (OMat24) \cite{barrosoluque2024openmaterials2024omat24} & 3D inorganic compounds & DFT (PBE(+U)/PAW) & 95,023,626 \\
Open Molecules 2025 (OMol25) \cite{levine2025omol25} & Organic molecules          & DFT ($\omega$B97M-V/def2-TZVPD) & 104,428,301 \\
OMol25 neutral singlet subset \cite{levine2025omol25} & Organic molecules           & DFT ($\omega$B97M-V/def2-TZVPD) & 34,363,525 \\
OMol25 non-neutral/non-singlet subset \cite{levine2025omol25} & Organic molecules    & DFT ($\omega$B97M-V/def2-TZVPD) & 42,769,648 \\
Open Polymers 2026 (OPoly26) \cite{levine2025opoly26} & Organic polymers          & DFT ($\omega$B97M-V/def2-TZVPD) & 6,318,494 \\
\midrule
\textcolor{black}{\textbf{Cumulative branch records}} & & & \textcolor{black}{\textbf{544,339,063}} \\
\textcolor{black}{\textbf{Unique structures}} & & & \textcolor{black}{\textbf{467,205,890}} \\
\bottomrule
\end{tabularx}
\par\vspace{2pt}\footnotesize\textcolor{black}{\emph{Note:} The two OMol25 subset branches are fully contained in the complete OMol25 branch. Their 77,133,173 records are therefore counted twice in the cumulative branch total, but only once in the unique-structure total. No other dataset overlaps are present.}
\end{table}

The pre-processed versions of all 16 datasets used for training are publicly available in ADIOS2 format through the OLCF Data Constellation release associated with this work~\cite{osti_2562660}.

\FloatBarrier

\subsection{Multi-Task HydraGNN Architecture}

\begin{figure*}[t]
\centering
\includegraphics[width=\textwidth]{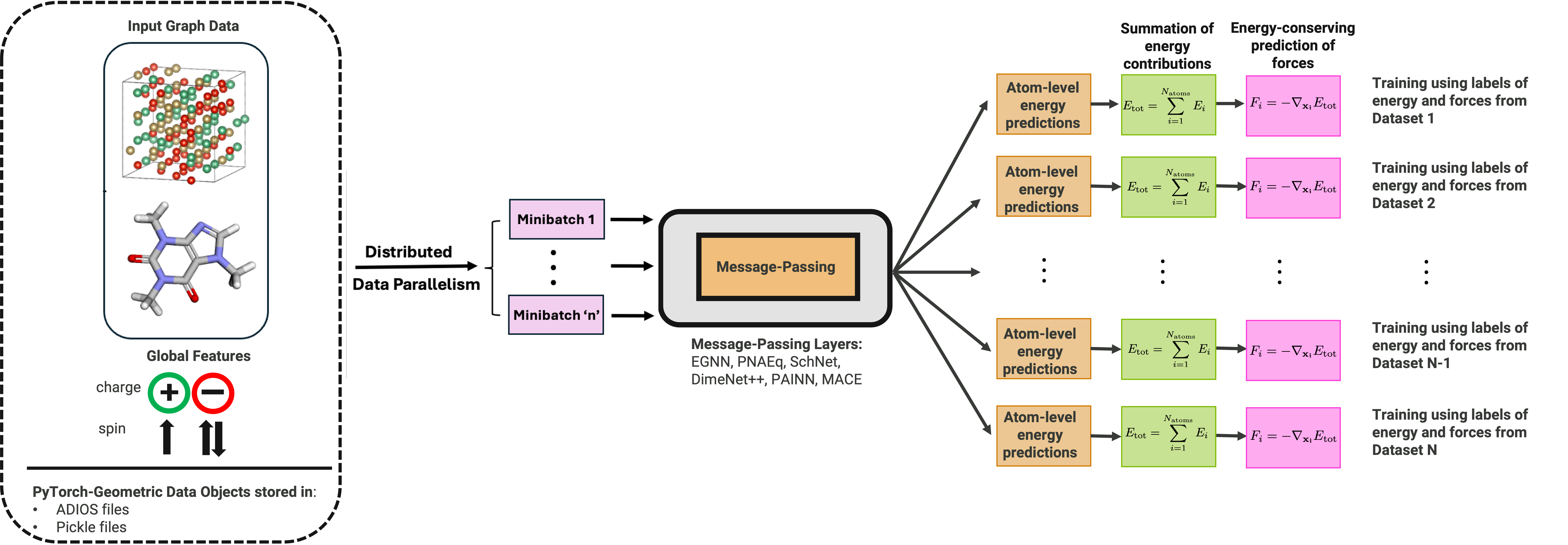}
\caption{Multi-task learning training for MLIPs on multi-source, multi-fidelity data, with shared message-passing representations and dataset-specific supervision heads. Message passing layers learn fundamental features that describe basic physics of interatomic interactions common to all datasets. 
Output heads customize features to the physics captured by the approximation theory used to generate a specific dataset}
\label{fig:mtl_mlip_workflow}
\end{figure*}

\textcolor{black}{The MTL design employed here is an established architectural choice motivated by our prior controlled comparisons of single-source, merged single-stream, and dataset-specific multi-task training (see Section~2). We extend this approach to 16 heterogeneous dataset branches and evaluate its training, adaptation, and deployment at exascale.} As shown in Fig.~\ref{fig:mtl_mlip_workflow}, shared message-passing layers learn transferable atomistic interaction features across datasets, while task-specific output heads \textcolor{black}{learn source-dependent energy and force labels without requiring identical energy references across datasets}. Compared with chemistry-routed MoE implemented in the UMA model, the head-level specialization implemented in HydraGNN directly targets cross-dataset heterogeneity during training: \textcolor{black}{it is intended to mitigate competing source-dependent supervision}, improves tolerance to source-specific reference/label offsets, simplifies handling of partially labeled batches, and supports modular onboarding of new datasets/fidelity levels. This parameter-sharing and task-decoupling strategy is designed to stabilize convergence when jointly fitting energies and derivative-based force labels at scale. \textcolor{black}{Here, multi-fidelity refers to training on data generated using different electronic-structure methods and computational settings, while preserving their respective energy references.}

\subsection{Scalable Data Movement and Distributed Training}

ADIOS2-backed loaders and lightweight Python wrappers orchestrate efficient parallel ingestion from the file system, while DDStore stages frequently accessed training shards closer to compute. Together, these mechanisms reduce repeated remote reads, lower inter-node data-transfer latency, and avoid overloading the shared file system during synchronized multi-node epochs \cite{adiosSoftwareX,ddstoreChoi}. This integration is central because it sustains the data pipeline needed for 2,048-node continual training and full-campaign execution at 16,384 GPUs, making multi-source atomistic GFM training practical under real file-system and memory constraints.

On Frontier, job scripts additionally exploit node-local non-volatile memory (NVMe) storage to stage both the Python virtual environment and training dataset files via multi-threaded \texttt{sbcast}. Environment tarballs and dataset bundles are broadcast to each node's burst buffer at job start, avoiding per-node pulls from the shared Lustre file system. For datasets exceeding a configurable size threshold, the NVMe stage is skipped and data is read directly from the parallel file system, balancing startup latency against local-storage capacity.

HydraGNN extends training parallelism with sharded distributed execution, distributed data parallelism, and optional task-parallel/device-mesh paths, significantly reducing per-device memory pressure and improving scalability. When task parallelism is enabled, proportional MPI rank allocation assigns compute in proportion to dataset size, avoiding idle ranks on smaller branches \cite{hydragnn4,LupoPasini2025HydraGNN,zhao2023fsdp}.

\subsection{Architecture Search and Resource-Aware Selection}

HPO is executed at large scale on Frontier to explore candidate configurations under a consistent training protocol while recording both predictive and systems metrics. The search space spans six equivariant message-passing neural network (MPNN) backbones: EGNN \cite{satorras2021egnn}, SchNet \cite{schutt2017schnet}, DimeNet \cite{gasteiger2020dimenet}, MACE \cite{batatia2022mace}, PaiNN \cite{schutt2021painn}, and our equivariant version of PNA \cite{corso2020pna} that we call PNAEq. Each MPNN backbone is paired with architecture-specific hyperparameter ranges for network depth, hidden-channel width, and interaction cutoff. \textcolor{black}{Architecture selection follows two stages. First, we rank completed HPO trials by their final validation loss and retain the most promising configurations. We then continue training the selected configurations on the full corpus using the same number of nodes and fixed-duration allocation windows. The second stage reveals how much training each architecture can complete within the available computational budget.} 

Table~\ref{tab:hpo_common_ranges} summarizes the hyperparameters tuned in every campaign. Among them, the message-passing hidden dimension was the only one whose range varied by backbone, because the computational and memory costs of increasing hidden dimension scale differently across message-passing implementations. Table~\ref{tab:hpo_model_specific} lists the additional backbone-specific hyperparameters (e.g., radial and spherical basis counts, maximum angular-momentum order) exposed for each MPNN family.

\begin{table}[t]
\caption{Hyperparameters tuned in all single-model HPO campaigns. Convolution depth and hidden dimensions were sampled as integers, force weight and learning rate were sampled as a continuous real-valued variable.}
\label{tab:hpo_common_ranges}
\centering
\scriptsize
\begin{tabularx}{\columnwidth}{@{}l X@{}}
\toprule
Hyperparameter & Range \\
\midrule
Convolution layers & 2--6 \\
Message-passing hidden dimension & EGNN/SchNet: 100--3000; DimeNet: 10--100; MACE/PaiNN/PNAEq: 100--1000 \\
Prediction-head layers & 2--4 \\
Prediction-head hidden dimension & 300--2000 \\
Force weight & $10$--$100$ \\
Learning rate & $10^{-5}$--$3\times10^{-3}$ \\
\bottomrule
\end{tabularx}
\end{table}

\begin{table}[t]
\caption{Additional HPO dimensions exposed for each MPNN family in the local SC26 HPO driver.}
\label{tab:hpo_model_specific}
\centering
\scriptsize
\begin{tabularx}{\columnwidth}{@{}l X@{}}
\toprule
MPNN & Additional tuned hyperparameters \\
\midrule
EGNN & None beyond the shared search space. \\
SchNet & Interaction filters: 6--300; Gaussian basis functions: 3--100. \\
DimeNet & Basis embedding size: 8--200; envelope exponent: 4--6; interaction embedding size: 16--500; output embedding size: 24--4000; layers after skip: 0--2; layers before skip: 0--2; radial basis count: 3--12; spherical basis count: 2--8. \\
MACE & Radial basis type: \{Bessel, Gaussian, Chebyshev\}; distance transform: \{Agnesi, Soft\}; radial basis count: 3--12; max $\ell$: 2--3; node max $\ell$: 1--3; correlation order: 2--3. \\
PaiNN & \textcolor{black}{Radial basis count: 3--12.} \\
PNAEq & Radial basis count: 3--12. \\
\bottomrule
\end{tabularx}
\end{table}

\textcolor{black}{For the selected PaiNN implementation, \texttt{hidden\_dim} specifies the target scalar/vector channel width (337): the first message/update block processes a one-channel scalar input and expands its representation to 337 channels, while the second block processes 337-channel features. The \texttt{num\_radial} setting selects four sinc-type distance functions. Their filter coefficients are generated internally, with three coefficients per input channel; the first and second blocks use filter maps of dimensions $4\!\to\!3$ and $4\!\to\!1{,}011$, respectively. The generic \texttt{num\_filters}=126 field is used by SchNet, not by PaiNN. The checkpoint-level architecture and parameter breakdown are documented in the Supporting Information.}\par

\subsection{Composition-Conditioned Branch Aggregation}

\textcolor{black}{The 16 training datasets have partially overlapping elemental and chemical coverage, but differ in electronic-structure settings and in the sizes of their atomistic configurations. This motivates learning how to combine source-specific prediction heads rather than prescribing one branch for each chemical family.} We implement a compact multilayer perceptron (MLP) \textcolor{black}{for composition-conditioned aggregation} of the 16 decoding branches of the MTL architecture. At inference, all branch decoders produce independent energy and force predictions for a given structure, and the MLP maps the structure's chemical-composition vector to softmax weights that blend these branch outputs into a single prediction. The learned composition-conditioned routing (see Section~\ref{branch_mlp_performance}) is a deployment mechanism for the multi-task model: rather than requiring a fixed branch assignment or heuristic selection, it enables the model to \textcolor{black}{combine source-specific predictions for compositions not clearly assigned to one training domain. This provides a composition-dependent prediction across the available source-specific heads.}

\textcolor{black}{Let $\mathbf R$ denote nuclear coordinates, $\mathbf Z$ atomic species, $E_k(\mathbf R,\mathbf Z)$ the energy from branch $k$, and $w_k(\mathbf Z)$ its nonnegative, softmax-normalized composition-dependent weight ($\sum_{k=1}^{16}w_k=1$). The combined energy is}
\begin{equation}
{\color{black} E_{\mathrm{mix}}(\mathbf R,\mathbf Z)=\sum_{k=1}^{16}w_k(\mathbf Z)E_k(\mathbf R,\mathbf Z).}
\label{energy_sum}
\end{equation}
\textcolor{black}{For fixed composition, the weights are independent of nuclear positions. Thus, provided that the same complete set of branch contributions is retained in both energy and force evaluation, the combined predictor satisfies}
\begin{equation}
{\color{black} \mathbf F_{\mathrm{mix}}=-\nabla_{\mathbf R}E_{\mathrm{mix}}=\sum_{k=1}^{16}w_k(\mathbf Z)\mathbf F_k,\qquad\mathbf F_k=-\nabla_{\mathbf R}E_k.}
\end{equation}
\textcolor{black}{The composition-conditioned sum \eqref{energy_sum} defines an effective potential whose forces are the negative coordinate gradient of its energy. The energy references of the individual electronic-structure sources remain source-specific.}

\subsection{Instrumentation and Precision-Aware Evaluation}

We instrument the full application with Omnistat \cite{omnistat2024}-based monitoring to capture GPU and central processing unit (CPU) utilization, runtime behavior, and computational resource usage during both HPO trials and production-scale training runs. These measurements make performance claims reproducible and allow direct attribution of efficiency gains to software and systems choices, not only model architecture.

We perform a systematic characterization of precision sensitivity at both inference and fine-tuning stages using the lead checkpoint identified by the large-scale continual-training campaign \cite{micikevicius2018mixed}. In contrast to conventional practice in atomistic materials modeling, where precision is often fixed a priori or evaluated only at the kernel level, we quantify the impact of BF16, FP32, and FP64 at the level of full-application predictions, including both energy and force outputs. This enables a direct assessment of how reduced precision affects deployment-time, accuracy, stability, and throughput in realistic workflows.

\section{Experimental Design}

\subsection{Applications and Workloads}
Performance is measured with GPTL, an HPC-friendly timing library, on the full HydraGNN application path, not on isolated kernels. Each run includes ADIOS2-based data ingestion, DDStore staging, distributed training, validation, and checkpoint/final artifact writes 
\cite{adiosSoftwareX,ddstoreChoi,hydragnn4}. We report wall-clock time-to-solution for complete training jobs and, where appropriate, decomposed phase times (data load, forward, backward, optimization) from the same end-to-end executions (Figure \ref{fig:scaling}).

Workloads are selected to reflect the study objective: training GFMs on realistic, heterogeneous materials and molecular datasets used in the study. The protocol includes six backbone-specific HPO campaigns, a fixed-budget 2,048-node continual-training stage for four selected configurations, and controlled experiments for scaling, downstream adaptation, inference optimization, and precision sensitivity. Fine-tuning, full-scale inference, and precision-sensitivity measurements use the lead checkpoint selected from the continual-training stage.

All reported results are measured, not projected. Strong-scaling and weak-scaling experiments for training are conducted across supercomputing facilities using consistent model configurations per experiment family, with only the intended scale variable changed (problem size for weak scaling, resources for strong scaling). For each configuration, multiple repeated runs are collected, and we report the central tendency with run-to-run variability. Numerical settings are held constant within each comparison, and any mixed-precision usage is explicitly stated in the corresponding result narrative so that quality-versus-performance tradeoffs are interpretable.

\subsection{Systems and Experimental Environment}
Scalable HPO runs, full scalable training of the selected HPO trials, and full scale inference are performed on OLCF Frontier, whereas strong and weak scaling experiments for training are executed on OLCF Frontier, ALCF Aurora, and NERSC Perlmutter with distributed multi-node allocations representative of production operations. Reproducibility is maintained through script-generated software environments provided in \path{HydraGNN/installation_DOE_supercomputers}\label{sec:install_scripts}, with platform-specific installation workflows. Across all three systems, the campaign uses Python~3.11 environments, PyTorch/PyG stacks appropriate to the local accelerator platform, ADIOS2 and DDStore for data movement, GPTL4Py for MPI-enabled multi-process timing measurement, and DeepHyper for HPO support. On Frontier, the stack is ROCm-based; on Aurora, it is built around the facility-provided XPU frameworks module; on Perlmutter, it targets CUDA/A100 execution. Measurement emphasizes whole-application behavior: runtime is recorded from job start to final reusable outputs, and Omnistat traces are synchronized with application logs to attribute bottlenecks to data movement, communication, and compute phases. To ensure full reproducibility, each reported point is tied to a concrete configuration tuple: model variant, dataset mix, precision mode, parallelism strategy, and resource count.

\section{Results}

All performance metrics in this section are produced from runs executed in the script-generated software environments described in Section~\ref{sec:install_scripts}. 

\subsection{Scalable HPO with Omnistat Telemetry}
We performed six large-scale HPO campaigns on OLCF Frontier, one for each message-passing neural network (MPNN) layer considered in this study: EGNN, SchNet, DimeNet, MACE, PaiNN, and PNAEq. A single HPO run used 1,024 Frontier nodes (8,192 GPUs) for 6 wall-clock hours.  In total, we conducted 10 HPO runs, with 1–3 runs allocated per MPNN model. DeepHyper explored 574 hyperparameter configurations, resulting in 186 successful trials that completed 20 training epochs. Omnistat instrumentation was enabled during these searches to capture system telemetry together with trial outcomes, allowing us to analyze HPO efficiency jointly with GPU utilization and runtime behavior.

The HPO workflow used the DeepHyper driver, with one MPNN fixed per run and a shared FP64 training protocol. Each HPO trial used 256 Frontier nodes, so four trials were evaluated concurrently within each 1,024-node run. New trials were proposed and launched as soon as running trials completed. To keep each trial within the 6-hour, 1,024-node HPO budget, every trial was evaluated on 10\% of each dataset rather than the full multi-source corpus. The search driver allows up to 200 evaluations per run and tunes a common set of architectural and optimization hyperparameters in all campaigns, while exposing additional MPNN-specific parameters only for the corresponding model family.

To address the wide variation in dataset size across our 16-source corpus, HydraGNN supports synthetic oversampling of smaller datasets, which we used during the HPO runs. When oversampling is enabled, minority datasets are resampled to a configurable target sample count, ensuring that smaller but scientifically valuable sources contribute meaningfully to every training epoch rather than being statistically drowned out by larger corpora.

\begin{figure}[t]
\centering
\includegraphics[width=\columnwidth]{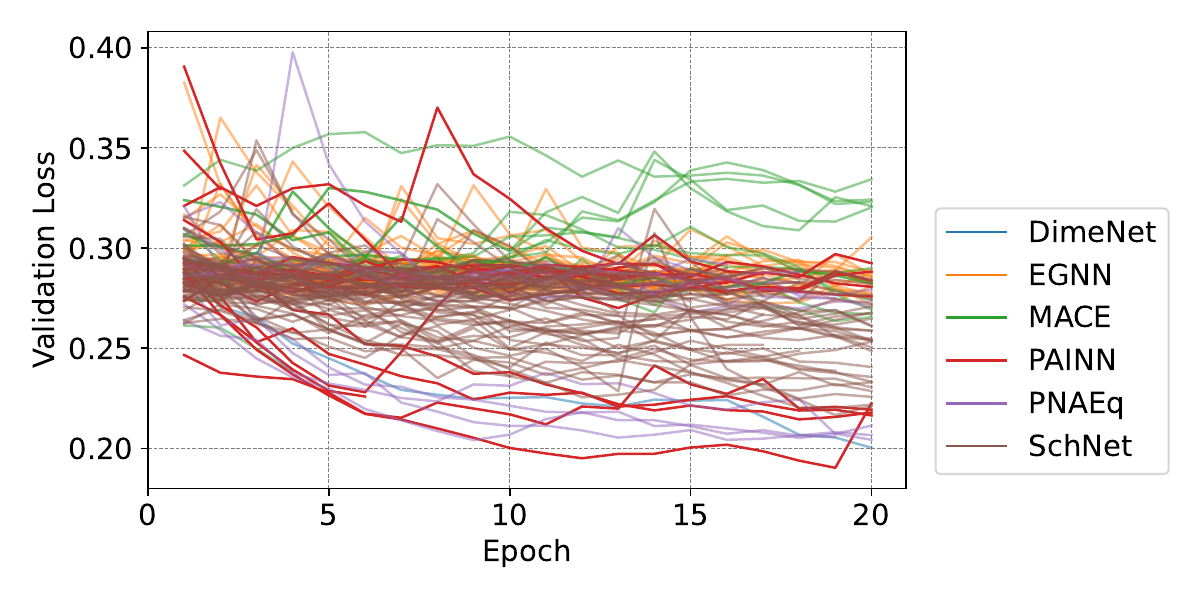}
\caption{Validation-loss trajectories for all trials from the six Frontier HPO campaigns, with one campaign per MPNN backbone. It shows 186 successful DeepHyper trials out of 574 explored configurations across MPNN backbones.}
\label{fig:hpo_loss_curves}
\end{figure}

The HPO search spaces are detailed in Tables~\ref{tab:hpo_common_ranges}--\ref{tab:hpo_model_specific} in the Innovations section. Hidden-dimension ranges vary by backbone to respect a common Frontier memory budget, so results should be interpreted as comparisons across architecture-specific feasible search spaces under fixed resources rather than as a perfectly matched hidden-dimension sweep. EGNN and SchNet are the lightest backbones computationally, MACE/PaiNN/PNAEq are intermediate due to their equivariant tensor operations, and DimeNet is the most memory-intensive because of its angular-basis triplet interactions and auxiliary embedding paths. Figure~\ref{fig:hpo_loss_curves} shows the validation-loss trajectories observed across all trials from the six backbone-specific HPO campaigns. 

Performing HPO at scale automates the identification of highly performing HydraGNN architectures, which otherwise would be unfeasible.

\subsection{Production-Scale Continual Training and Lead-Model Selection}
\textcolor{black}{We ranked completed HPO trials by their validation loss at the final epoch. The configurations selected for continual training achieved losses of 0.200350 (DimeNet), 0.204303 (PNAEq), 0.216497 (PaiNN), and 0.221185 (SchNet). All evaluated EGNN and MACE configurations had validation losses above 0.27, with little evidence of improvement over the observed epochs (Fig.~\ref{fig:hpo_loss_curves}), and were not promoted. DimeNet and PNAEq therefore had lower validation losses than PaiNN at the end of HPO.} We promoted these four models to continual training runs at 2,048 nodes (16,384 GPUs) on Frontier to test whether they could achieve convergence at production scale. The selected configurations were evaluated under matched 2,048-node, fixed-duration allocation windows with checkpoint/restart. \textcolor{black}{At this stage, we assessed how far each architecture could progress within the same wall-clock budget, rather than relying solely on its HPO validation loss.} This fixed-budget protocol exposed a strong architecture-dependent cost to make epoch progress at scale and provided a clear basis for model ranking under realistic campaign constraints. From this study, the lead model (a PaiNN backbone) advanced to 100 training epochs and delivered the predictive quality used for downstream analysis. 

Under identical production budgets, PaiNN was the only promoted backbone to reach 100 training epochs within the fixed 2,048-node production runs, \textcolor{black}{and was therefore selected as the lead model for downstream experiments. This selection uses the number of epochs completed within the fixed production budget.} PaiNN’s pairwise interaction structure with low-order scalar/vector updates reduces per-layer geometric work, memory traffic, and synchronization pressure, enabling more optimizer steps and more completed epochs within the same 12-hour production windows. Under our fixed-budget continual-training protocol, the lower time-to-epoch translated directly into a practical convergence advantage.

Table~\ref{tab:lead_model} summarizes the configuration of the lead PaiNN model. \textcolor{black}{The complete saved HydraGNN checkpoint contains 12,205,540 parameters, including its shared encoder and all 16 dataset-specific heads; the FP64 parameter tensors alone occupy approximately 93\,MiB. The previously reported count covered 12,090,960 parameters and omitted a lazily instantiated graph-conditioning projector.} This moderate footprint supports direct deployment in simulation and screening workflows that require high inference throughput. Consequently, the downstream fine-tuning study, full-scale inference campaign, and precision-sensitivity analysis reported in this paper all use this mature reference lead PaiNN checkpoint.

\begin{table}[t]
\caption{Lead PaiNN model configuration selected from the HPO campaign and used for all downstream analysis.}
\label{tab:lead_model}
\centering
\scriptsize
\begin{tabularx}{\columnwidth}{@{}l X@{}}
\toprule
\textbf{Parameter} & \textbf{Value} \\
\midrule
Convolution layers & 2 \\
\textcolor{black}{Scalar/vector hidden channels} & \textcolor{black}{337} \\

\textcolor{black}{Radial basis functions} & \textcolor{black}{4 (sinc-type)} \\
Cutoff radius & 5\,\AA \\
Max neighbors & 20 \\
Output branches & 16 (graph-level) \\
\textcolor{black}{Head architecture} & \textcolor{black}{Each: $337\to50\to50\to776\to776\to1$; unshared} \\
Optimizer & AdamW, lr $= 6.34\times10^{-4}$ \\
Batch size & 25 per GPU \\
Loss & MAE \\
Force weight & $\approx$ 94.8 \\
Precision & FP64 \\
\textcolor{black}{Parameters / FP64 weight memory} & \textcolor{black}{12,205,540 / ${\sim}$93\,MiB} \\
Epochs completed & 100 \\
\bottomrule
\end{tabularx}
\end{table}

\textcolor{black}{The checkpoint contains two PaiNN message/update blocks. The first processes the scalar atomic-number input with one feature channel and expands its scalar and vector representation to 337 channels; the second operates on 337-channel scalar/vector features. Four radial functions and a cosine cutoff at 5~\AA{} define the distance dependence, with no independently configured PaiNN interaction-filter width. Sixteen dataset-specific prediction networks map the shared graph representation to source-specific scalar energies. Checkpoint inspection verified 12,205,540 HydraGNN parameters, including 1,601,460 shared-encoder and 10,604,080 prediction-head parameters. A separate 24,528-parameter BranchWeightMLP supplies the aggregation weights at inference. The Supporting Information reports the detailed architecture and reconciles the full checkpoint count with the optimizer-covered count of 12,090,960.}\par

The trained weights of this lead PaiNN model, together with checkpoints from the production-scale model-selection campaign, are publicly released through the OLCF Data Constellation~\cite{osti_2562660}.

\subsection{Cross-System Strong and Weak Scaling}
\begin{figure}
    \centering
    \includegraphics[width=1.0\columnwidth]{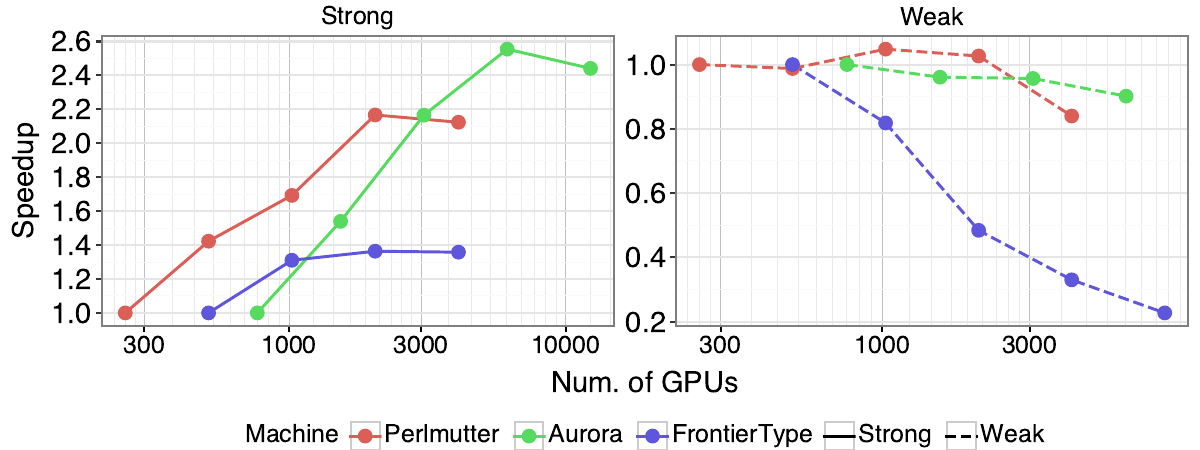}
    \includegraphics[width=1.0\columnwidth]{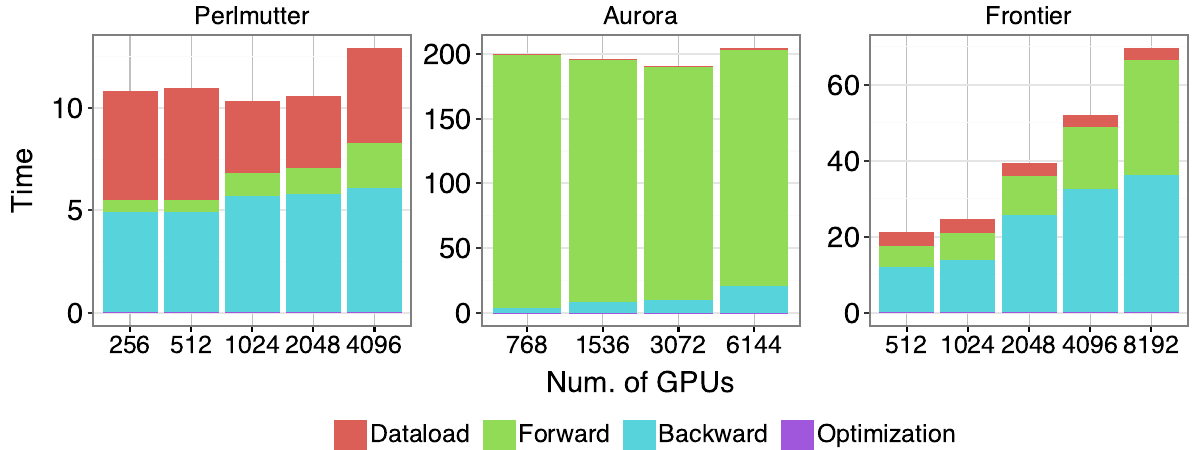}
\caption{Strong and weak scaling results, along with time decomposition, are presented for Perlmutter, Aurora, and Frontier. Speedup (y-axis) is normalized to the 32-node performance on each system, with the number of GPUs shown on the x-axis. The decomposed timing corresponds to the weak scaling results, detailing time spent on data loading, forward pass, backward pass, and optimization. }
    \label{fig:scaling}
\end{figure}

Matched strong- and weak-scaling results are presented for the lead reference checkpoint to isolate platform portability from architecture-dependent training differences. We keep the model configuration fixed and compare OLCF Frontier, ALCF Aurora, and NERSC Perlmutter under consistent problem definitions, so efficiency and time-to-solution differences reflect platform and software behavior. 	
Figure~\ref{fig:scaling} shows the normalized runtime for both strong and weak scaling on Perlmutter, Aurora, and Frontier. 
For strong scaling, we fix the total number of samples (102,400) and evenly distribute them across multiple GPUs. For weak scaling, the number of samples per GPU is kept constant as the system scales. 
We observe near-linear strong scaling up to 2,048 GPUs on Perlmutter, 6,144 GPUs on Aurora, and 1,024 GPUs on Frontier, showing portable scaling across three distinct leadership-class systems under consistent problem requirements. The performance drop at the largest scales in strong scaling is attributed to network saturation during gradient aggregation, which is a common bottleneck.

For weak scaling, each GPU processes a constant workload while gradient aggregation is performed across all GPUs. Because Frontier delivers the highest per-GPU computational efficiency, the fixed cost of distributed gradient synchronization becomes a larger fraction of runtime at smaller node counts than on Aurora or Perlmutter, leading to an earlier and more visible decline in efficiency. In contrast, Aurora and Perlmutter remain flatter over the same range because computation continues to dominate communication. Overall, degradation is gradual on all systems.

The bottom panels of Figure~\ref{fig:scaling} provide a time decomposition for the weak scaling results, illustrating the contributions from data loading, forward pass, backward pass, and optimization. Gradient synchronization time is included within the backward pass and cannot be separated explicitly due to its highly asynchronous nature. On Frontier, the backward phase takes a substantial portion of the total runtime and grows with node count, whereas it remains a minor contribution on Aurora and Perlmutter. Taken together, these results show that the workflow is operationally stable and scalable across three distinct leadership-class accelerator architectures.

\FloatBarrier
\subsection{Downstream Transfer Characterization with Representative State-of-the-Art Models}
\label{sec:finetuning}

All downstream HydraGNN experiments use the lead PaiNN checkpoint emerging from the fixed-budget exascale training campaign. We evaluate the checkpoint in two complementary ways. First, HydraGNN-only ablations compare training from scratch, frozen-backbone adaptation, and unfrozen full-model adaptation across a broad downstream task suite. Second, we compare HydraGNN with representative pretrained atomistic models: UMA \cite{wood2025uma}, MACE-OFF23 \cite{kovacs2025maceoff}, MACE-MH-1 \cite{batatia_cross_2025}, and MACE-MP-0 \cite{batatia2024foundation}. \textcolor{black}{These checkpoints span different training corpora, architectures, and target conventions.} The benchmark includes zero-shot inference and all adaptation modes available from our runs---frozen-backbone fine-tuning, full fine-tuning, and low-rank adaptation (LoRA). The open-source scripts used to generate the zero-shot and fine-tuning comparisons are available in the \texttt{uma-sota-comparison} branch of the HydraGNN GFM fine-tuning repository.\footnote{\url{https://github.com/allaffa/HydraGNN_GFM_FineTuning4Materials/tree/uma-sota-comparison}}. 

\textcolor{black}{UMA and MACE were fine-tuned using separate, architecture-specific protocols. UMA used Adam with a learning rate of $10^{-4}$: QM9 runs used 10 epochs and a batch size of 8, while MD17 used a batch size of 4 and ran for 4 epochs (full fine-tuning) or 20 epochs (frozen backbone and LoRA). MACE used its native training workflow, with batch sizes of 16 on QM9 and 4 on MD17. The standard configurations specified 50 epochs at $10^{-3}$ for full fine-tuning and 10 epochs at $5\times10^{-3}$ for LoRA. QM9 used an energy-only objective; MD17 used energies and forces.}

\textcolor{black}{Energy errors were evaluated using dataset-specific reference conventions. Zero-shot and fine-tuned UMA and MACE predictions were corrected using elemental energy offsets fitted exclusively on the training split and then applied without refitting to evaluation predictions. MD17 energy predictions were mean-centered before computing errors.}

The HydraGNN downstream suite spans small molecules, inorganic crystals, and condensed-phase trajectories (Table~\ref{tab:finetuning_datasets}). Potential-energy-surface (PES) tasks include energy and/or force prediction, whereas MB-jdft2d and MB-is\_metal probe transfer to targets that are not direct energy-and-force objectives. This distinction is useful because it separates representation transfer from simple reuse of the pretraining target.

\begin{table*}[t]
\caption{Downstream task suite used to evaluate transfer of the lead HydraGNN checkpoint. Size indicates the number of structures available in the HydraGNN fine-tuning protocol.}
\label{tab:finetuning_datasets}
\centering
\small
\renewcommand{\arraystretch}{1.12}
\begin{tabularx}{\textwidth}{@{}l r l X@{}}
\toprule
\textbf{Dataset} & \textbf{Size} & \textbf{Domain} & \textbf{Target(s)} \\
\midrule
QM9 \cite{ramakrishnan2014qm9} & ${\sim}$130k & Small organic molecules & Atomization energy \\
MD17 \cite{chmiela2017md17} & ${\sim}$134k & Uracil ab-initio MD trajectory & Total energy and forces \\
Wiggle150 \cite{brew2025wiggle150} & 150 & Strained molecular conformers & Relative conformer energy \\
MS25--MgO-2$\times$2 \cite{ms25benchmark} & 10k & MgO(100) oxide surface & Energy and forces \\
MS25--MgO-4$\times$4 \cite{ms25benchmark} & 10k & MgO(100) oxide surface & Energy and forces \\
MS25--H$_2$O-64 \cite{ms25benchmark} & -- & Condensed-phase water & Energy and forces \\
MS25--H$_2$O-192 \cite{ms25benchmark} & -- & Condensed-phase water & Energy and forces \\
MS25--CHA \cite{ms25benchmark} & -- & Zeolite framework & Energy and forces \\
MS25--Reaction \cite{ms25benchmark} & 10k & CH$_4$ dissociation on Pt(111) & Energy and forces \\
MS25--HEA \cite{ms25benchmark} & 10k & FeNiCrCoCu high-entropy alloy & Energy and forces \\
MS25--Zr--O \cite{ms25benchmark} & 10k & Amorphous Zr--O oxides & Energy and forces \\
OQMD \cite{saal2013oqmd} & ${\sim}$16k & ABX$_3$ inorganic perovskite subset & Formation energy \\
ABX$_3$ (\texttt{abc3}) \cite{chenebuah2023abc3} & ${\sim}$4.5k & Inorganic perovskites & Formation energy \\
MB-jdft2d \cite{dunn2020benchmarking} & 636 & Layered inorganic materials & Exfoliation energy \\
MB-is-metal \cite{dunn2020benchmarking} & ${\sim}$106k & Inorganic crystals & Metal/non-metal classification \\
\bottomrule
\end{tabularx}
\end{table*}

\paragraph{Molecular and conformational transfer characterization.}
\textcolor{black}{Tables~\ref{tab:qm9_sota_main}--\ref{tab:wiggle_sota_main} compare HydraGNN with pretrained reference models across downstream chemistries, targets, and adaptation strategies.} The answer changes with chemistry, target, data regime, and model history. \textcolor{black}{HydraGNN supports frozen-backbone and full-model reuse of its shared pretrained representation.} To make errors directly readable across result tables, energy errors are reported in eV (or eV/atom when normalized per atom) and force errors in eV/\AA{}. Values below $10^{-2}$ are written in scientific notation, while larger values use conventional decimal notation, generally to about three significant figures.

On QM9 (Table~\ref{tab:qm9_sota_main}), frozen-backbone transfer preserves most of the value of the exascale-trained HydraGNN representation: HydraGNN reaches $0.0127$~eV/atom, improving over HydraGNN training from scratch ($0.0150$~eV/atom) and remaining close to the strongest zero-shot reference points, $\num{9.90e-3}$--$0.0104$~eV/atom for UMA, MACE-OFF23 (medium), and MACE-MH-1 (OMol). This is a useful operating regime for a checkpoint trained across a much broader heterogeneous corpus, because adaptation requires updating only the downstream head. \textcolor{black}{For the evaluated UMA and MACE configurations, fine-tuning gives higher QM9 test errors than zero-shot prediction:} MACE-MH-1 rises to $0.0338$--$0.0470$~eV/atom, MACE-OFF23 to $0.0798$--$0.128$~eV/atom, and UMA to $0.210$--$0.721$~eV/atom. QM9 therefore shows that lightweight reuse of HydraGNN can be competitive even when specialized zero-shot checkpoints retain a modest accuracy advantage.

\begin{table}[t]
\centering
\caption{QM9 atomization-energy held-out test MAE (eV/atom). Lower is better. Values below $10^{-2}$~eV are shown in scientific notation; larger values use conventional decimal notation, generally to about three significant figures. The MACE model names identify the evaluated checkpoint variants. ``--'' denotes an unavailable result.}
\label{tab:qm9_sota_main}
\small
\begin{tabularx}{0.84\linewidth}{@{}Xr@{}}
\toprule
Model / strategy & MAE \\
\midrule
UMA zero-shot & \textbf{$\num{9.90e-3}$} \\
MACE-OFF23 (medium) zero-shot & 0.0101 \\
MACE-MH-1 (OMol) zero-shot & 0.0104 \\
HydraGNN GFM frozen FT & 0.0127 \\
HydraGNN scratch & 0.0150 \\
HydraGNN GFM unfrozen FT & 0.0233 \\
MACE-MP-0 (medium) zero-shot & 0.0280 \\
MACE-MH-1 (OMol) full FT (50 epochs) & 0.0338 \\
MACE-MH-1 (OMol) LoRA FT & 0.0470 \\
MACE-OFF23 (medium) LoRA FT & 0.0798 \\
MACE-OFF23 (medium) full FT (50 epochs) & 0.128 \\
UMA frozen FT & 0.210 \\
UMA LoRA FT & 0.375 \\
UMA full FT & 0.721 \\
\bottomrule
\end{tabularx}
\end{table}

MD17 uracil (Table~\ref{tab:md17_sota_main}) provides a boundary case for HydraGNN transfer. Here the HydraGNN pretrained initialization does not improve on scratch training: scratch reaches 0.113~eV energy MAE and 0.491~eV/\AA{} force MAE, while frozen and unfrozen fine-tuning remain in the same error range. The molecular zero-shot references are stronger, with UMA at 0.064~eV and 0.299~eV/\AA{} and MACE-OFF23 at 0.068~eV and 0.312~eV/\AA{}. At the same time, aggressive adaptation is not automatically beneficial for those checkpoints: full fine-tuning of MACE-MH-1 (OMol) and MACE-OFF23 gives energy errors above 4~eV and force errors above 17~eV/\AA{}, while UMA adaptation produces energy errors above 5~eV and force errors above 23~eV/\AA{}. Thus, MD17 delineates a regime in which HydraGNN pretraining offers limited transfer benefit and also reinforces the need to choose adaptation protocols conservatively.

\begin{table}[t]
\centering
\caption{MD17 uracil held-out test MAE for energy (eV) and forces (eV/\AA{}). Lower is better. The table includes the newly completed MACE-MH-1 (OMol) full-fine-tuning result. ``--'' denotes an unavailable result.}
\label{tab:md17_sota_main}
\small
\begin{tabularx}{0.90\linewidth}{@{}Xrr@{}}
\toprule
Model / strategy & Energy & Forces \\
\midrule
UMA zero-shot & \textbf{0.064} & \textbf{0.299} \\
MACE-OFF23 (medium) zero-shot & 0.068 & 0.312 \\
HydraGNN scratch & 0.113 & 0.491 \\
HydraGNN ANI-1x recycled head & 0.165 & 0.748 \\
HydraGNN GFM frozen FT & 0.175 & 0.737 \\
HydraGNN GFM unfrozen FT & 0.188 & 0.794 \\
MACE-MH-1 (OMol) full FT & 4.08 & 17.4 \\
MACE-OFF23 (medium) LoRA FT & 4.11 & 17.5 \\
MACE-OFF23 (medium) full FT & 4.15 & 17.5 \\
UMA LoRA FT & 5.05 & 23.7 \\
UMA full FT & 5.07 & 24.2 \\
UMA frozen FT & 5.53 & 23.7 \\
\bottomrule
\end{tabularx}
\end{table}

Wiggle150 (Table~\ref{tab:wiggle_sota_main}) is the clearest illustration of the intended reuse value of the multi-source HydraGNN representation in an extreme low-data regime. With only 150 downstream structures, unfrozen HydraGNN fine-tuning reaches 0.679~eV, followed by frozen HydraGNN at 0.773~eV and training from scratch at 0.870~eV. The best non-HydraGNN adapted result is MACE-MH-1 with LoRA at 2.02~eV, followed by full MACE-MH-1 fine-tuning at 2.22~eV and full UMA fine-tuning at 2.39~eV. The updated MACE zero-shot sweep gives 2.41~eV for MACE-OFF23 (medium), 2.44~eV for MACE-MH-1 (OMol), and 2.88~eV for MACE-MP-0 (medium). HydraGNN unfrozen fine-tuning therefore reduces MAE by approximately $3.0\times$ relative to the best adapted external reference and by approximately $3.5\times$ relative to the best external zero-shot result. This result is important for the HydraGNN design objective: broad pretraining provides a useful initialization precisely where generating additional first-principles labels is most constrained.

\begin{table}[t]
\centering
\caption{Wiggle150 total-energy held-out test MAE (eV). Lower is better. The MACE zero-shot values are updated from the expanded variant sweep. ``--'' denotes an unavailable result.}
\label{tab:wiggle_sota_main}
\small
\begin{tabularx}{0.84\linewidth}{@{}Xr@{}}
\toprule
Model / strategy & MAE \\
\midrule
HydraGNN GFM unfrozen FT (BF16) & \textbf{0.679} \\
HydraGNN GFM frozen FT (BF16) & 0.773 \\
HydraGNN scratch (BF16) & 0.870 \\
MACE-MH-1 (OMol) LoRA FT & 2.02 \\
MACE-MH-1 (OMol) full FT & 2.22 \\
UMA full FT & 2.39 \\
MACE-OFF23 (medium) zero-shot & 2.41 \\
MACE-MH-1 (OMol) zero-shot & 2.44 \\
UMA zero-shot & 2.47 \\
UMA frozen FT & 2.53 \\
UMA LoRA FT & 2.55 \\
MACE-MP-0 (medium) zero-shot & 2.88 \\
MACE-OFF23 (medium) LoRA FT & 3.32 \\
MACE-OFF23 (medium) full FT & 5.39 \\
\bottomrule
\end{tabularx}
\end{table}

\paragraph{Inorganic and condensed-phase transfer characterization.}
On ABX$_3$ (Table~\ref{tab:abx3_sota_main}), which is named \texttt{abc3} in the open-source benchmark repository and chemically corresponds to ABX$_3$-type perovskite structures, HydraGNN full-model, energy-only fine-tuning for 100 epochs reaches 0.240~eV/atom. In this inorganic task, the external reference points are stronger: MACE-MP-0 (medium) improves from 0.118~eV/atom zero-shot to 0.0361~eV/atom after full fine-tuning and 0.0530~eV/atom with LoRA, while UMA gives 0.118~eV/atom in evaluation-calibrated zero-shot testing and 5.876~eV/atom after two epochs of full fine-tuning with direct, uncorrected evaluation. ABX$_3$ therefore serves as a useful boundary condition for the broad HydraGNN checkpoint: transfer is not uniformly advantageous across all target domains, and domain/pretraining alignment can outweigh the benefit of a more heterogeneous pretraining corpus. This result motivates retaining multiple adaptation modes rather than treating full-model fine-tuning as the default.

\begin{table}[t]
\centering
\caption{ABX$_3$ formation-energy held-out test MAE (eV/atom). Lower is better. The dataset is named \texttt{abc3} in the benchmark repository. HydraGNN uses energy-only full-model fine-tuning for 100 epochs from the pretrained checkpoint. The updated MACE-MP-0 (medium) full-fine-tuning result is 0.0361~eV/atom after 100 epochs.}
\label{tab:abx3_sota_main}
\small
\begin{tabular}{lr}
\toprule
Model / strategy & MAE \\
\midrule
MACE-MP-0 (medium) full FT & \textbf{0.0361} \\
MACE-MP-0 (medium) LoRA FT & 0.0530 \\
MACE-MP-0 (medium) zero-shot & 0.118 \\
UMA zero-shot (OMat task) & 0.118 \\
HydraGNN GFM full FT (100 epochs) & 0.240 \\
UMA full FT & 5.876 \\
\bottomrule
\end{tabular}
\end{table}

MS25 is used primarily as an adaptation stress test because it spans eight chemically and structurally distinct condensed-phase systems while allowing the same three HydraGNN fine-tuning strategies to be exercised throughout. Table~\ref{tab:ms25_unified} places these results beside zero-shot UMA and MACE-MP-0 references and attempted external full-model fine-tuning. A central HydraGNN result is adaptation coverage: all three HydraGNN strategies complete on all eight systems, providing a consistent downstream pathway across oxides, water, zeolites, alloys, and reactive surfaces. Under the imposed two-hour external fine-tuning limit, UMA full-model fine-tuning does not complete on any of the eight systems and MACE-MP-0 completes on MgO-2$\times$2, Reaction, and Zr--O but not on the other five. Missing fine-tuning results are therefore reported as dashes rather than being omitted.

Within HydraGNN, MS25 illustrates why configurable adaptation is useful. Full-model fine-tuning gives the lowest HydraGNN energy MAE on MgO-2$\times$2, MgO-4$\times$4, CHA, HEA, and Zr--O, whereas freezing the decoder and adapting the message-passing representation is best on H$_2$O-64, H$_2$O-192, and Reaction. Freezing the message-passing layers is not the lowest-energy-error choice, but it gives the lowest HydraGNN force MAE on several systems. The energy and force rankings therefore select different parameter-update strategies. Rather than identifying a single preferred fine-tuning recipe, these results support HydraGNN's configurable adaptation design: the useful degree of parameter reuse depends on both the downstream system and the target quantity.

The external checkpoints provide complementary reference regimes. UMA and MACE-MP-0 are often strong zero-shot predictors on MS25, whereas MACE-MP-0 full fine-tuning is exceptionally accurate on the three systems for which it completes: energy MAEs of $\num{5.20e-5}$, $\num{7.04e-4}$, and $0.0195$~eV/atom for MgO-2$\times$2, Reaction, and Zr--O, with force MAEs of $\num{6e-3}$, $0.023$, and $0.106$~eV/\AA{}. At the same time, five MACE-MP-0 full-fine-tuning runs and all eight UMA full-fine-tuning runs exceed the two-hour limit. MS25 therefore separates three quantities that a pure accuracy leaderboard would conflate: zero-shot predictive quality, the gain obtainable from adaptation, and adaptation time-to-solution. \textcolor{black}{HydraGNN completes the tested adaptation strategies across all eight systems, enabling direct comparison of parameter-reuse strategies across the suite.} \textcolor{black}{UMA and MACE energies are mean-centered per system for MS25 evaluation, including the fine-tuned pathways implemented in the benchmark scripts.} Force MAEs use eV/\AA{} throughout.

\begin{table*}[t]
\centering
\caption{Unified MS25 comparison across HydraGNN, UMA, and MACE-MP-0. Panel (a) reports energy MAE in eV/atom for every method and strategy. The MACE-MP-0 and UMA zero-shot values are converted from the benchmark-reported meV/atom values; zero-shot energies are mean-centered per system before error evaluation. Values below $10^{-2}$ are shown in scientific notation; larger values use conventional decimal notation, generally to about three significant figures. Panel (b) reports force MAE in eV/\AA{} for all methods. ``--'' denotes a fine-tuning run that did not complete within the two-hour training limit, so no converged result is reported. Lower is better.}
\label{tab:ms25_unified}
\scriptsize
\setlength{\tabcolsep}{3.3pt}
\renewcommand{\arraystretch}{1.12}
\resizebox{\textwidth}{!}{%
\begin{tabular}{lrrrrrrr}
\toprule
\multicolumn{8}{l}{\textit{(a) Energy MAE (eV/atom)}} \\
\addlinespace[2pt]
System & \multicolumn{3}{c}{HydraGNN fine-tuning (eV/atom)} & \multicolumn{2}{c}{UMA} & \multicolumn{2}{c}{MACE-MP-0} \\
\cmidrule(lr){2-4}\cmidrule(lr){5-6}\cmidrule(lr){7-8}
 & Full & Frozen MP & Frozen decoder & Zero-shot & Full FT & Zero-shot & Full FT \\
\midrule
MgO-2$\times$2 & $\num{5.04e-3}$ & 0.0139 & $\num{5.10e-3}$ & $\num{7.23e-4}$ & -- & $\num{1.16e-3}$ & $\num{5.20e-5}$ \\
MgO-4$\times$4 & $\num{2.35e-3}$ & $\num{2.65e-3}$ & $\num{2.35e-3}$ & $\num{2.64e-4}$ & -- & $\num{4.50e-4}$ & -- \\
H$_2$O-64 & $\num{5.08e-3}$ & 0.301 & $\num{3.54e-3}$ & $\num{2.12e-3}$ & -- & $\num{2.41e-3}$ & -- \\
H$_2$O-192 & $\num{3.45e-3}$ & 0.230 & $\num{2.44e-3}$ & $\num{1.76e-3}$ & -- & $\num{2.00e-3}$ & -- \\
CHA & 0.0162 & 0.371 & 0.0166 & $\num{1.92e-3}$ & -- & $\num{1.08e-3}$ & -- \\
HEA & 0.0201 & 0.0304 & 0.0205 & 0.102 & -- & 0.0977 & -- \\
Reaction & $\num{3.97e-3}$ & 0.0195 & $\num{3.02e-3}$ & $\num{3.33e-3}$ & -- & $\num{4.06e-3}$ & $\num{7.04e-4}$ \\
Zr--O & 0.0990 & 0.276 & 0.111 & $\num{4.97e-3}$ & -- & 0.0462 & 0.0195 \\
\midrule
\multicolumn{8}{l}{\textit{(b) Force MAE (eV/\AA{})}} \\
\addlinespace[2pt]
System & \multicolumn{3}{c}{HydraGNN fine-tuning} & \multicolumn{2}{c}{UMA} & \multicolumn{2}{c}{MACE-MP-0} \\
\cmidrule(lr){2-4}\cmidrule(lr){5-6}\cmidrule(lr){7-8}
 & Full & Frozen MP & Frozen decoder & Zero-shot & Full FT & Zero-shot & Full FT \\
\midrule
MgO-2$\times$2 & 0.474 & 0.417 & 0.491 & 0.024 & -- & 0.064 & $\num{6e-3}$ \\
MgO-4$\times$4 & 0.370 & 0.399 & 0.371 & 0.021 & -- & 0.061 & -- \\
H$_2$O-64 & 0.475 & 3.199 & 0.472 & 0.134 & -- & 0.147 & -- \\
H$_2$O-192 & 0.471 & 3.415 & 0.471 & 0.134 & -- & 0.147 & -- \\
CHA & 1.976 & 18.799 & 2.239 & 0.083 & -- & 0.101 & -- \\
HEA & 0.794 & 0.654 & 1.036 & 0.281 & -- & 0.319 & -- \\
Reaction & 0.514 & 0.400 & 0.546 & 0.073 & -- & 0.127 & 0.023 \\
Zr--O & 1.605 & 0.330 & 1.499 & 0.061 & -- & 0.219 & 0.106 \\
\bottomrule
\end{tabular}%
}
\end{table*}

\paragraph{Implications for adaptation.}
\textcolor{black}{The downstream results show where pretraining helps and where it does not. On Wiggle150, fine-tuning the pretrained HydraGNN model reduces energy MAE from 0.870~eV (training from scratch) to 0.679~eV using only 150 structures. On MD17 uracil, training from scratch instead gives lower energy and force errors (0.113~eV and 0.491~eV/\AA{}) than either pretrained HydraGNN adaptation strategy. On ABX$_3$, the domain-aligned MACE-MP-0 model achieves 0.0361~eV/atom after fine-tuning, compared with 0.240~eV/atom for HydraGNN. These results reveal substantial differences in the benefits of pretraining across downstream chemistries and prediction targets, highlighting the importance of accounting for task-specific characteristics when selecting pretrained models and adaptation strategies.} On QM9, frozen HydraGNN transfer is competitive without updating the backbone. Across the eight MS25 systems, HydraGNN completes all tested adaptation strategies, while the preferred strategy varies with the target and system.

\begin{figure}[t]
\centering
\includegraphics[width=0.7\textwidth]{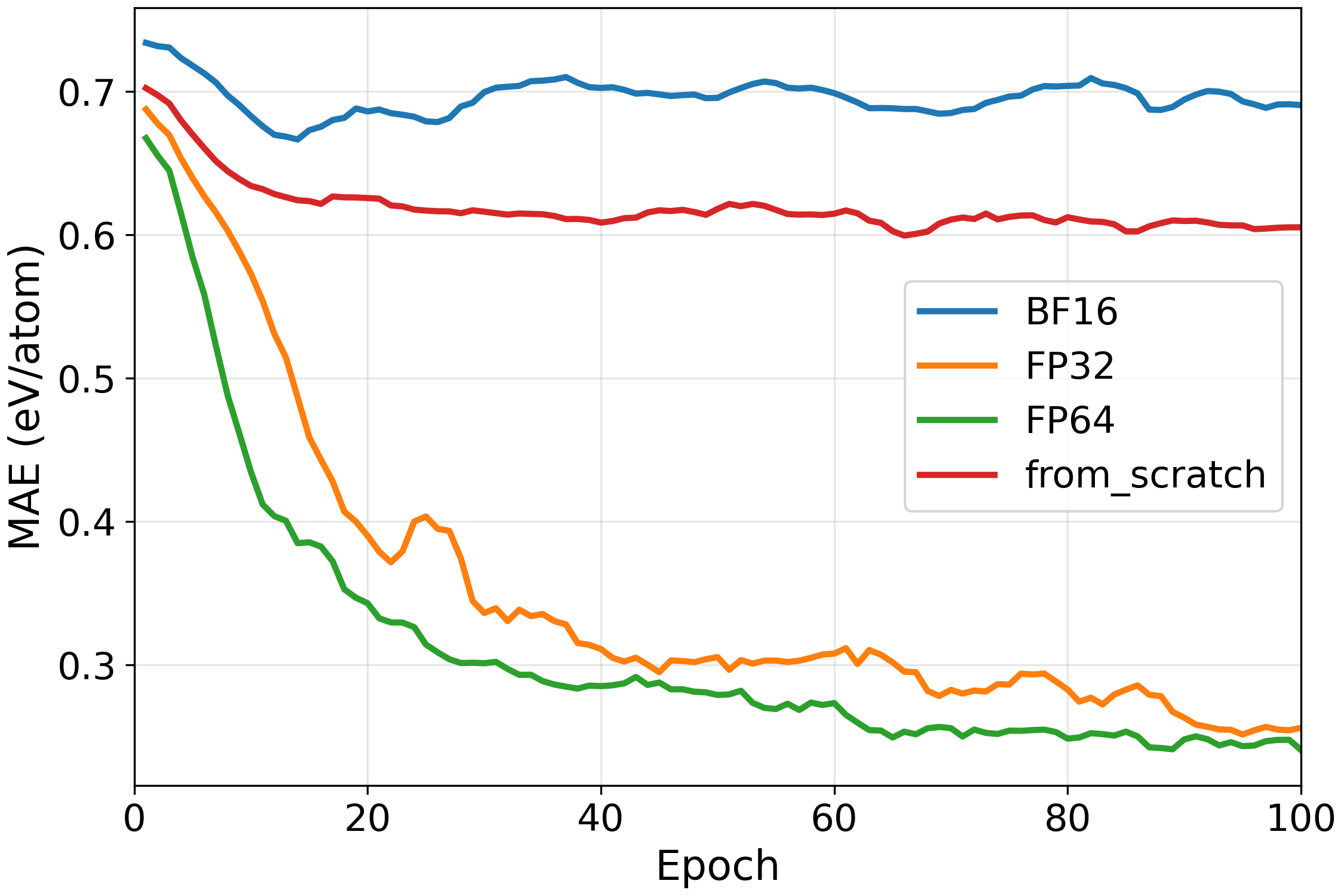}
\caption{Validation-loss trajectories for HydraGNN fine-tuning on the OQMD dataset. A smoothing window of 5 epochs was used.}
\label{fig:oqmd_valCurves}
\end{figure}

\paragraph{HydraGNN precision sensitivity and non-PES transfer.}
The broader HydraGNN-only experiments provide two additional diagnostics that are not covered by the direct UMA/MACE comparison. First, OQMD fine-tuning is sensitive to numerical precision. FP64 unfrozen fine-tuning reaches 0.227~eV/atom, compared with 0.245~eV/atom in FP32 and 0.684~eV/atom in BF16, while FP64 training from scratch gives 0.596~eV/atom (Table~\ref{tab:oqmd_results} and Fig.~\ref{fig:oqmd_valCurves}). Second, the Matbench tasks show weaker transfer when the downstream objective is not directly aligned with the PES pretraining target. For MB-jdft2d, scratch training gives the lowest mean MAE, while on MB-is\_metal the frozen and unfrozen pretrained backbones provide only modest gains over scratch. These results reinforce the conclusion that pretraining scale alone does not guarantee transfer when the target observable differs from energy and force prediction.

\begin{table}[t]
\centering
\caption{HydraGNN OQMD validation MAE (eV/atom) after 100 epochs, illustrating precision sensitivity during full fine-tuning.}
\label{tab:oqmd_results}
\begin{tabular}{lc}
\toprule
Method & OQMD MAE \\
\midrule
FP64 training from scratch & 0.596 \\
BF16 unfrozen backbone & 0.684 \\
FP32 unfrozen backbone & 0.245 \\
FP64 unfrozen backbone & \textbf{0.227} \\
\bottomrule
\end{tabular}
\end{table}

\begin{table}[H]
\centering
\caption{HydraGNN non-PES Matbench results. MB-jdft2d energy errors are in eV/atom; MB-is\_metal reports classification metrics. Energy values are reported to approximately three significant figures.}
\label{tab:matbench_nonpes}
\small
\begin{tabular}{llrrr}
\toprule
Dataset & Metric & Scratch & Frozen & Unfrozen \\
\midrule
MB-jdft2d & Avg. MAE & \textbf{0.0478} & 0.0507 & 0.0483 \\
 & Std. MAE & 0.0152 & 0.0158 & 0.0144 \\
 & Avg. RMSE & 0.117 & 0.117 & \textbf{0.113} \\
 & Max. error & 1.56 & \textbf{1.53} & 1.54 \\
\midrule
MB-is\_metal & ROCAUC & 0.911 & 0.924 & \textbf{0.924} \\
 & Std. ROCAUC & 0.002 & 0.003 & 0.003 \\
 & F1 & 0.809 & 0.823 & \textbf{0.825} \\
 & Accuracy & 0.835 & 0.847 & 0.848 \\
\bottomrule
\end{tabular}
\end{table}

\FloatBarrier
\subsection{Branch Weight Learning via MLP}
\label{branch_mlp_performance}
The branch-weighting MLP \textcolor{black}{comprises 24,528 parameters} and maps per-element chemical-composition vectors to per-branch blending weights. It was trained on 49 million samples uniformly distributed across all training datasets, minimizing a combined MSE loss over energies and forces computed from the softmax-normalized weighted average of the branch predictions. The teacher signal used one-hot assignment, directing each structure's target weights toward the branch corresponding to its source dataset. Training was performed across 64 GPUs of the Frontier supercomputer with an AdamW optimizer initialized at a learning rate of $1\times10^{-4}$ and a ReduceLROnPlateau scheduler decayed the rate by a factor of 0.5 after 10 epochs, reaching a final learning rate of $6.25\times10^{-6}$ after 207 epochs with a validation loss of $3.31\times10^{-3}$. 
\textcolor{black}{The resulting aggregation mechanism permits inference without assigning each structure to a single training dataset. The weights are learned from source identity and energy--force prediction objectives.}

\textcolor{black}{We assessed the trained BranchWeightMLP through two sequential, controlled composition-probe experiments. First, to isolate the effect of elemental composition at fixed system size, we generated 195 distinct compositions in 13 synthetic chemical families (15 per family), each with exactly 100 atoms. Element counts were encoded as the same 118-dimensional, unnormalized atomic-number histogram used to train the MLP. The synthetic configurations had deterministic simple-cubic coordinates, which do not enter the weighting network. We evaluated the frozen MLP in FP64, converted its 16 logits to softmax weights, and collected the complete weight vector for each composition to construct the routing heatmap. No HydraGNN backbone evaluation or electronic-structure calculation was required.}\par
\textcolor{black}{Second, after constructing the heatmap, we held stoichiometric ratios fixed and varied the absolute elemental counts for five compositions: Nb--Ta (10, 20, 50, 100, and 200 atoms), TiO$_2$ (6, 30, 60, and 120), CH$_4$ (5, 20, 50, and 100), LiFePO$_4$ (7, 14, 35, and 70), and Nb--Ta--V--Hf--Zr (10, 20, 50, 100, and 200). This follow-up test isolates size-dependent routing because the MLP takes absolute elemental counts, not normalized atomic fractions, as input. For every probe we recorded the complete 16-branch distribution, the highest-weight branch, and its Jensen--Shannon divergence from the smallest-size case. These tests characterize learned branch selection, not energy--force prediction accuracy.}\par

\begin{figure}[H]
\centering
\includegraphics[width=0.7\textwidth]{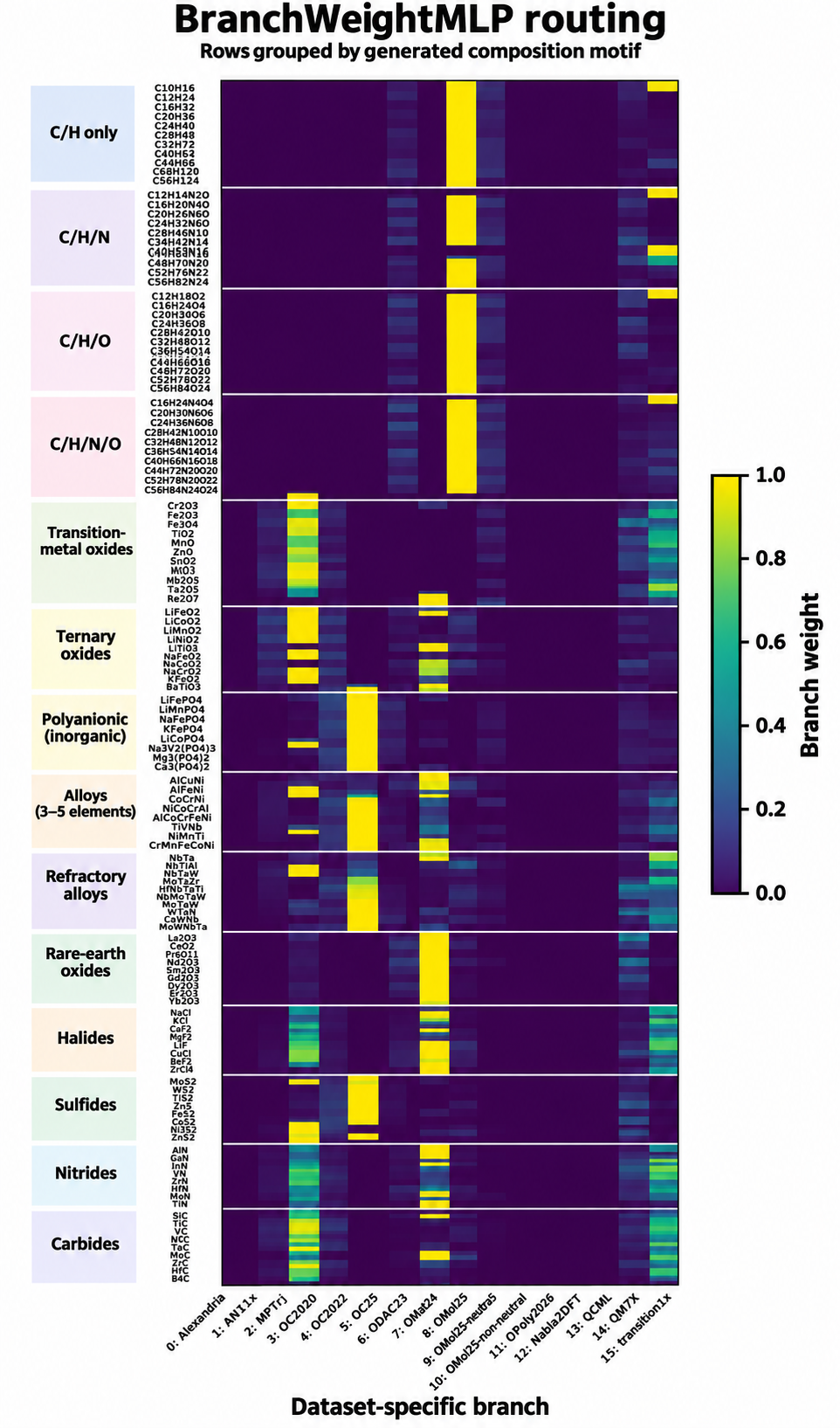}
\caption{\textcolor{black}{BranchWeightMLP routing weights for representative composition probes from the fixed-size experiment. Each composition contains 100 atoms and is encoded by its 118-dimensional elemental-count vector. Rows denote compositions grouped by generated chemical family; columns denote the 16 dataset-specific prediction branches. Color indicates the softmax weight assigned to each branch. The plot characterizes the weighting network directly, without evaluating the HydraGNN backbone.}}
\label{fig:branch_weight_routing}
\end{figure}

\textcolor{black}{Figure~\ref{fig:branch_weight_routing} shows composition-dependent preferences among the 16 prediction heads. The fixed-size probes predominantly associate C--H-rich compositions with OMol25, transition-metal oxides with OC2022, and several alloy families with MPTrj or transition1x. The average largest branch weight is 0.898 across the 195 probes, indicating that the learned distribution is often concentrated. These patterns illustrate a data-driven combination of heads from partially overlapping chemical domains, without a hard-coded assignment of each composition to one dataset.}\par

\begin{table}[t]
\centering
\small
\caption{\textcolor{black}{Dominant prediction branch along the separate, fixed-stoichiometry system-size probes. Branch sequences follow the order of atom counts within each row. The branches represent learned weighting preferences, not hard-coded chemical assignments.}}
\label{tab:branch_mlp_size_transitions}
\begin{tabularx}{\textwidth}{@{}p{0.20\textwidth}p{0.23\textwidth}X@{}}
\toprule
\textbf{Composition} & \textbf{Number of atoms} & \textbf{Dominant branch, in size order} \\
\midrule
Nb--Ta & 10, 20, 50, 100, 200 & OMat24 $\rightarrow$ MPTrj $\rightarrow$ MPTrj $\rightarrow$ transition1x $\rightarrow$ transition1x \\
TiO$_2$ & 6, 30, 60, 120 & MPTrj $\rightarrow$ OC2022 $\rightarrow$ OC2022 $\rightarrow$ OC2022 \\
CH$_4$ & 5, 20, 50, 100 & OMol25 $\rightarrow$ OMol25 $\rightarrow$ OMol25 $\rightarrow$ OMol25 \\
LiFePO$_4$ & 7, 14, 35, 70 & OMat24 $\rightarrow$ MPTrj $\rightarrow$ MPTrj $\rightarrow$ MPTrj \\
Nb--Ta--V--Hf--Zr & 10, 20, 50, 100, 200 & OMat24 $\rightarrow$ OC2022 $\rightarrow$ MPTrj $\rightarrow$ OMat24 $\rightarrow$ OMat24 \\
\bottomrule
\end{tabularx}
\end{table}

\textcolor{black}{Table~\ref{tab:branch_mlp_size_transitions} reports the follow-up size-sensitivity experiment, in which elemental fractions were held fixed while the number of atoms increased. Although the dominant branch sometimes changes, most transitions occur among heads trained on datasets with broadly compatible chemical coverage: Nb--Ta alternates among OMat24, MPTrj, and transition1x; TiO$_2$ and LiFePO$_4$ move between inorganic-materials branches; and CH$_4$ consistently favors OMol25. The Nb--Ta--V--Hf--Zr series temporarily selects OC2022, a less intuitive match for a bulk refractory alloy, but otherwise favors OMat24 and MPTrj. The results illustrate how the learned weights respond to both chemical identity and structure size while generally favoring chemically plausible sources, without requiring a prescribed assignment to a single branch.}\par

\subsection{Optimizing the Inference Pipeline}
\label{single-node-optimization-inference}

\begin{figure}[t]
  \centering
  \includegraphics[width=1.0\columnwidth]{%
    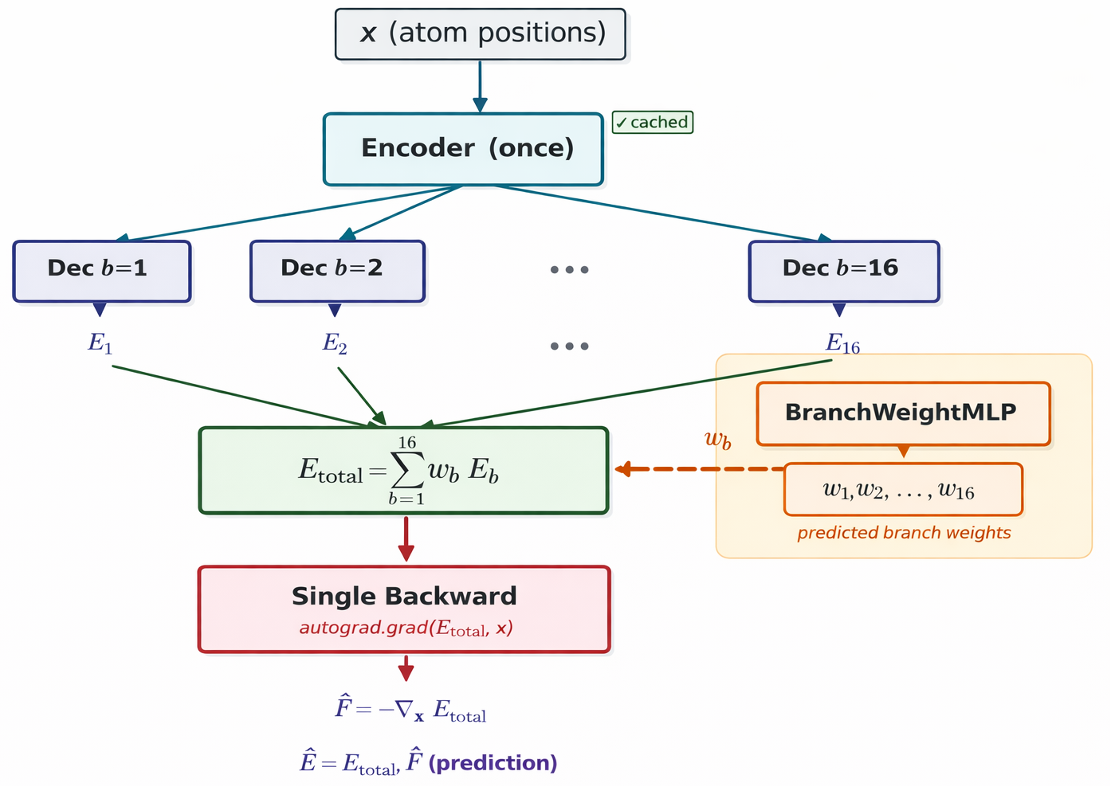}
\caption{Fused Gradient Inference Pipeline. The encoder executes once and its
    activations are cached.  All 16 decoder heads run forward-only, their
    energies are combined into a weighted sum, and a single backward pass
    computes forces.}
  \label{fig:fused-reuse}
\end{figure}

We present a systematic hill-climbing optimization of the inference for the multi-headed lead PaiNN model. Starting from a na\"{i}ve baseline that
executes 16 encoder-decoder-backward passes per batch, we
progressively eliminate inefficiencies through four complementary optimization techniques. (1) The optimization that we refer to as \textit{encoder reuse} caches the shared-trunk computation and runs it once instead of sixteen
times, yielding a consistent $1.6\times$ speedup across all precisions and GPU
families. (2) The optimization called \textit{branch skipping} exploits the sparsity of the BranchWeightMLP's
softmax outputs: about 81\% of branches carry negligible weight for a
given structure. \textcolor{black}{In this exploratory variant, low-weight branches retain their energy contributions but do not contribute to the computed forces; consequently, this variant does not enforce the energy--force gradient relationship and was not used for the production-scale screening.} (3)
The optimization called \textit{fused gradient} (Figure~\ref{fig:fused-reuse}) combines all 16 decoder energies into a single weighted sum and computes forces with one
\texttt{autograd.grad} call, achieving an $11$--$14\times$ speedup. (4) Finally, \texttt{torch.compile}
enables kernel fusion in the encoder convolution blocks. \textcolor{black}{The sequential single-node benchmark reaches up to a $33\times$ speedup when branch skipping is included (Table~\ref{tab:throughput}); this is separate from the full-branch production inference path. With all branches retained, fused-gradient FP64 inference differentiates the complete weighted energy.} FP32
introduces a $|\Delta E|$ of 0.021\,eV in the reported precision comparison, while BF16 shows $|\Delta E|{\approx}0.080$\,eV. For production deployment, we choose FP64 on Frontier's MI250 GPUs to maintain the precision required for scientific inference.

Table~\ref{tab:throughput} reports per-GPU throughput for each optimization
step on MI250X (64\,GiB HBM) and MI350X (288\,GiB HBM) at
two precision levels; the bottom row gives the end-to-end speedup.  All
measurements use 40\,000 structures (atoms 2--500) on a single 8-GPU node.

\begin{table}[htbp]
  \centering
  \caption{Per-GPU Throughput (structures/s). \textcolor{black}{The branch-skipping variants omit low-weight force contributions and are separate from the full-branch production screening path.}}
  \label{tab:throughput}
  \small
  \begin{tabular}{l rr rr rr}
    \toprule
    & \multicolumn{2}{c}{FP64}
    & \multicolumn{2}{c}{FP32} \\
    \cmidrule(lr){2-3}\cmidrule(lr){4-5}
    Optimization
      & {MI250} & {MI350}
      & {MI250} & {MI350} \\
    \midrule
    Baseline
      & 23    & 106
      & 42    & 190   \\
    $+$\,Encoder\ reuse
      & 35    & 169
      & 66    & 313  \\
    $+$\,Branch skip
      & 165   & 743
      & 308   & 1250 \\
    $+$\,Fused gradient
      & 306   & 1428
      & 602   & 2134 \\
    $+$\,\texttt{torch.compile}
      & 738   & 3012
      & 1405  & 6050 \\
    \cmidrule(lr){2-3}\cmidrule(lr){4-5}\cmidrule(lr){6-7}
    Speedup
      & 33$\times$  & 28$\times$
      & 33$\times$  & 32$\times$ \\
    \bottomrule
  \end{tabular}
\end{table}

\textcolor{black}{We evaluated energy--force consistency on ten crystalline configurations against an independent FP64 reference that computes forces as the negative gradient of the complete 16-branch weighted energy. We compared three methods. \emph{Full-branch fused inference} retains all 16 branches in both the reported energy and its gradient. \emph{Force-only skipping} reports the same complete energy but omits the force contributions of branches whose softmax weights are below 0.2. In the force-only skipping approach, all 16 branches contribute to the reported energy, but only branches whose weights exceed the prescribed threshold contribute to the forces. Consequently, the computed forces are not, in general, the negative gradient of the reported energy, violating energy–force consistency. In contrast, energy–force truncation excludes the same low-weight branches from both the energy and force calculations, without renormalizing the retained weights. The resulting forces are therefore the negative gradient of the truncated energy for a fixed set of retained branches, although this energy defines an approximate potential distinct from the complete 16-branch model.}

\textcolor{black}{Table~\ref{tab:force_consistency_r1c6} compares each method's forces with the complete FP64 reference. Full-branch fused inference agrees to rounding precision. Force-only skipping and energy--force truncation have identical force errors because they retain exactly the same force contributions; the distinction is which energy those forces correspond to. With force-only skipping, the mean force RMSE was $5.99\times10^{-7}$\,eV/\AA{} and the largest component error was $1.53\times10^{-5}$\,eV/\AA{} (Al). On average, the threshold excluded 92.5\% of branch-force contributions but only 0.00491\% of total softmax weight. Direct differentiation and finite differences confirmed the mismatch for force-only skipping; when the same branches were omitted from the energy, the maximum direct-gradient residual decreased to $3.89\times10^{-15}$\,eV/\AA{} for fixed retained-branch sets. Thus, the small measured force error does not make force-only skipping conservative, nor does it bound its error on other structures. The earlier approximately 81\% skip fraction pertains to a separate throughput experiment. The production screening and structural relaxations used the complete 16-branch energy and forces without skipping.}\begin{table}[H]
\centering
\begingroup\color{black}
\caption{FP64 force discrepancies relative to the complete 16-branch energy-gradient reference, across ten crystalline configurations. Force-only skipping retains the full reported energy but omits selected force contributions; energy--force truncation omits the same branches from both energy and forces. Their errors relative to the full model are identical, but only truncation preserves the local energy--force relation. The threshold of 0.2 applies to this comparison.}
\label{tab:force_consistency_r1c6}
\small
\setlength{\tabcolsep}{3.5pt}
\begin{tabular}{@{}p{0.31\columnwidth}rrr@{}}
\toprule
Method & Mean RMSE & Max. RMSE & Max. component \\
 & \multicolumn{3}{c}{(eV/\AA)} \\
\midrule
Full-branch fused & $1.79\times10^{-16}$ & $1.00\times10^{-15}$ & $2.78\times10^{-15}$ \\
Force-only skipping & $5.99\times10^{-7}$ & $5.99\times10^{-6}$ & $1.53\times10^{-5}$ \\
Energy--force truncation & $5.99\times10^{-7}$ & $5.99\times10^{-6}$ & $1.53\times10^{-5}$ \\
\bottomrule
\end{tabular}
\endgroup
\end{table}

By treating numerical precision as a controllable design dimension rather than an implementation detail, this analysis establishes practical operating regimes for large-scale atomistic foundation models and clarifies the trade-offs required to achieve efficient scientific inference. \textcolor{black}{The single-node optimization study characterizes several performance options, while the production screening uses the complete 16-branch predictor.}

\subsection{Inference on Billions of Atomistic Structures on Frontier}
At near-full-machine scale on OLCF Frontier (9,300 compute nodes), we sustain end-to-end screening of 1.1 billion atomistic structures in 50 seconds using the full HydraGNN inference pipeline. \textcolor{black}{All 16 prediction branches were evaluated in this run; branch skipping was not enabled.} As an illustrative cost comparison, assuming 30 minutes for a representative first-principles calculation on a GPU-enabled node, the same number of evaluations would correspond to approximately 550 million GPU-hours; even at the same nominal GPU count, this estimate is roughly 6.7 years of continuous execution. The comparison is intended to indicate workload scale rather than claim equivalence between ML inference and a specific first-principles calculation.

Let $N_{\mathrm{struct}}$ denote the total number of screened structures, $T_{\mathrm{wall}}$ the measured wall-clock time, and $P$ the number of participating GPUs. We report sustained throughput as
\[
\mathrm{Throughput} = \frac{N_{\mathrm{struct}}}{T_{\mathrm{wall}}},
\quad
\mathrm{Per\mbox{-}GPU\ throughput} = \frac{N_{\mathrm{struct}}}{P\,T_{\mathrm{wall}}}.
\]
For the Frontier production run, these quantities evaluate to approximately 21.8 million structures/s overall, 293 structures/s/GPU, and 1.3 billion structures/min at system scale. Because the checkpoint was trained in FP64 and then evaluated under controlled BF16/FP32/FP64 inference settings, the same experiment also provides a deployment-time view of the accuracy--throughput tradeoff: lower precision increases screening throughput, but introduces measurable deviations, with FP32 yielding $|\Delta E| = 0.021$ eV and BF16 $|\Delta E| \approx 0.080$ eV.

\begingroup
\color{black}
\subsection{Numerical Precision in Coupled Atomic and Cell Relaxation}
\label{sec:precision_relaxation}
To evaluate the numerical robustness of the frozen multi-branch potential beyond single-point inference, we performed paired FP32 and FP64 structural relaxations for 13 periodic crystalline prototypes spanning elemental covalent solids, metals, binary compounds, and a perovskite. Each ASE-generated prototype was expanded into a $2\times2\times2$ supercell. We constructed compressed and expanded initial states by scaling the cell vectors by 0.95 and 1.05, respectively, and displacing each atom by 0.10--0.15\,\AA{} along a seeded random direction. The same perturbation was applied to the two strain states of each prototype, and each FP32/FP64 pair started from precisely the same geometry. This yielded 26 initial configurations and 52 relaxation runs (Table~\ref{tab:precision_relax_setup}).

The pretrained HydraGNN checkpoint (epoch 97) and BranchWeightMLP (epoch 207) remained frozen. At each step, the weighting MLP used the 118-element vector of absolute atomic counts, all 16 branch energies contributed to the weighted energy, and branch skipping was disabled. Atomic forces and cell stresses were obtained by differentiation of this complete energy. The periodic graph was rebuilt as positions and cell vectors evolved, using a 5.0\,\AA{} neighbor radius and at most 20 neighbors per atom. The model and weighting network were evaluated consistently in the designated precision; neutral, zero-spin inputs were used throughout.

\begin{table}[t]
\color{black}
\centering
\caption{Protocol for the paired FP32/FP64 relaxation benchmark. The generalized-force convergence test includes both atomic and scaled cell degrees of freedom.}
\label{tab:precision_relax_setup}
\small
\begin{tabularx}{\textwidth}{@{}p{0.33\textwidth}X@{}}
\toprule
Setting & Value \\
\midrule
Initial systems & 13 crystalline prototypes; $2\times2\times2$ supercells; 5\% compression/expansion; 0.10--0.15\,\AA{} seeded atomic displacements \\
Evaluation & Paired FP32 and FP64; 26 starts; 52 runs; same frozen 16-branch model and composition-weighting MLP \\
Energy, force, and stress & All 16 energy branches retained; differentiated weighted energy; branch skipping off \\
Optimizer & ASE FIRE with \texttt{UnitCellFilter}; full cell and atomic positions relaxed; no symmetry constraints or imposed external pressure \\
Stopping protocol & Maximum step 0.03\,\AA; generalized-force tolerance 0.05\,eV/\AA; at most 500 optimization steps; cell factor $N$ \\
Graph and composition & Rebuilt periodic graph, 5.0\,\AA{} cutoff, 20-neighbor cap; 118-dimensional absolute elemental-count vector \\
\bottomrule
\end{tabularx}
\end{table}

All 52 runs met the generalized-force stopping criterion, corresponding to all 26 starting configurations converging independently in both FP32 and FP64. The two compressed and expanded diamond-carbon precision pairs converged and exhibited maximum endpoint differences of $1.03\times10^{-6}$\,eV/atom in energy and $1.21\times10^{-5}$\,\AA{} in cell-vector length. The expanded ZnO and SrTiO$_3$ starting configurations already satisfied the optimizer criterion at step zero in both precisions and therefore underwent no geometry updates.

Table~\ref{tab:precision_relax_results} gives precision-pair comparisons by prototype. Across all 26 converged pairs, the largest difference in final predicted energy was $1.62\times10^{-4}$\,eV/atom (expanded MgO); the largest final cell-vector length difference was 0.0166\,\AA{} (also expanded MgO), and the largest cell-angle difference was 0.00552$^{\circ}$ (expanded Si). Most other pairs show substantially smaller differences, confirming consistent FP32 and FP64 numerical outcomes under repeated energy, force, and stress evaluations.

\begin{table}[t]
\color{black}
\centering
\caption{FP32 versus FP64 relaxation outcomes. Each prototype has a compressed and an expanded starting configuration; both precisions are counted as a converged pair only if both meet the optimizer's generalized-force stopping criterion. Energy and length columns show the maximum absolute FP32--FP64 endpoint difference across the two starting configurations for each prototype, with all pairs converged. These values measure precision agreement, not accuracy relative to electronic-structure references.}
\label{tab:precision_relax_results}
\small
\begin{tabular}{@{}lccc@{}}
\toprule
Prototype & Converged pairs & Max. $|\Delta E|$ (eV/atom) & Max. $|\Delta a_i|$ (\AA) \\
\midrule
C (diamond) & 2/2 & $1.03\times10^{-6}$ & $1.21\times10^{-5}$ \\
Si (diamond) & 2/2 & $1.45\times10^{-5}$ & $1.17\times10^{-4}$ \\
Ge (diamond) & 2/2 & $1.01\times10^{-6}$ & $1.29\times10^{-5}$ \\
Al (FCC) & 2/2 & $1.27\times10^{-6}$ & $1.06\times10^{-5}$ \\
Cu (FCC) & 2/2 & $5.91\times10^{-7}$ & $3.14\times10^{-6}$ \\
Ni (FCC) & 2/2 & $1.16\times10^{-6}$ & $1.35\times10^{-7}$ \\
Fe (BCC) & 2/2 & $4.37\times10^{-7}$ & $1.24\times10^{-7}$ \\
W (BCC) & 2/2 & $4.34\times10^{-7}$ & $1.36\times10^{-6}$ \\
MgO (rocksalt) & 2/2 & $1.62\times10^{-4}$ & $1.66\times10^{-2}$ \\
NaCl (rocksalt) & 2/2 & $6.16\times10^{-8}$ & $1.74\times10^{-7}$ \\
GaAs (zincblende) & 2/2 & $1.04\times10^{-6}$ & $9.37\times10^{-5}$ \\
ZnO (wurtzite) & 2/2 & $2.40\times10^{-7}$ & $6.42\times10^{-8}$ \\
SrTiO$_3$ (perovskite) & 2/2 & $2.67\times10^{-7}$ & $3.49\times10^{-8}$ \\
\midrule
Total & 26/26 & --- & --- \\
\bottomrule
\end{tabular}
\end{table}

These results establish precision consistency of the full-branch learned potential during coupled structural and cell relaxation. The reported stopping criterion refers to the maximum generalized force in \texttt{UnitCellFilter}, not a separate requirement of vanishing individual atomic forces and stress components. Likewise, agreement between FP32 and FP64 does not by itself establish recovery of a unique equilibrium geometry or agreement with first-principles reference calculations.
\endgroup

\section{Discussion}
The experiments show that efficient scaling atomistic foundation models requires simultaneous treatment of statistical imbalance, systems cost, and downstream adaptation behavior. Dataset-specific prediction heads prevent the largest sources from being the only supervision pathway into the shared representation, while oversampling during HPO ensures that smaller datasets remain visible to the optimizer.

\textcolor{black}{The downstream benchmarks assess whether pretraining on heterogeneous atomistic datasets improves predictions for new materials and tasks.} HydraGNN was designed to combine broad multi-source pretraining with configurable downstream reuse, and the benchmarks show both where that design is effective and where domain-aligned external checkpoints retain an advantage. Frozen HydraGNN transfer is competitive on QM9, full-model transfer is particularly effective in the 150-structure Wiggle150 regime, and all tested HydraGNN adaptation strategies execute across the heterogeneous MS25 suite. \textcolor{black}{For MD17 and ABX$_3$, training from scratch or using a model pretrained on more closely related data gives better predictions.} This variation is scientifically consistent with the premise of heterogeneous pretraining: transfer depends on the interaction between pretraining coverage and the downstream chemistry, target observable, data volume, reference convention, and optimization protocol. 

MS25 further highlights a systems property that conventional downstream leaderboards rarely capture: time-to-solution. HydraGNN completes all three tested adaptation strategies on all eight systems, enabling a full comparison of parameter-reuse choices across diverse condensed-phase chemistries. Under the two-hour external fine-tuning limit, UMA full fine-tuning does not complete on any of the eight systems and MACE-MP-0 completes on three; on those three completed cases, MACE-MP-0 achieves lower errors than the HydraGNN variants. \textcolor{black}{Thus, HydraGNN can be adapted to every system tested, whereas the specialized reference models achieve lower errors on the cases for which their fine-tuning completes.} This distinction is central for leadership-class scientific workflows, where a useful model must be judged by both predictive quality and the cost and reliability of specialization. \textcolor{black}{On MD17, fine-tuning UMA and MACE increases both energy and force errors relative to their zero-shot predictions.} The Matbench results provide an additional boundary condition: a non-PES target such as MB-jdft2d can remain competitive when trained from scratch, emphasizing that broad pretraining does not remove the need for objective alignment.

The architecture-selection results also show why predictive error alone is insufficient for leadership-scale training. Under a fixed compute budget, different equivariant backbones convert node-hours into optimizer steps at substantially different rates. PaiNN reached 100 epochs within the matched production-scale campaign and therefore provided the most mature checkpoint for downstream study. \textcolor{black}{Within the available computing allocation, PaiNN completed more training epochs than the alternative architectures.} \textcolor{black}{Its combination of validation performance and training throughput motivated its selection for the downstream experiments.} The more detailed runtime interpretation in the Supporting Information relates this behavior to pairwise versus angular interactions, tensor-product cost, memory traffic, and synchronization.

Cross-system experiments on Frontier, Aurora, and Perlmutter indicate that portability is achievable at the workflow level when data movement, environment construction, distributed execution, and measurement are treated as first-class components. Strong scaling remains close to linear over substantial ranges, while the weak-scaling decompositions expose the increasing contribution of gradient synchronization, particularly on Frontier where high per-GPU compute throughput makes communication visible earlier. These results suggest that future scientific-GFM optimization should target communication and backward-pass overlap at least as aggressively as forward-kernel throughput.

Deployment introduces a second optimization regime. The original multi-head inference path redundantly repeated shared encoder work and force differentiation for each branch. \textcolor{black}{Single-node experiments show that reusing the shared encoder, combining force differentiation, compiling the model, and optionally skipping low-weight branches can accelerate reliable inference by up to 33$\times$. The 1.1-billion-structure screening on 9,300 Frontier nodes used the full 16-branch energy and force calculation, without branch skipping. The ten-structure FP64 comparison (Table~\ref{tab:force_consistency_r1c6}) confirms full-branch fused force agreement to rounding precision. Force-only skipping keeps the complete energy but discards some force contributions, breaking the energy--force relation; energy--force truncation removes the same branches from both quantities and preserves that relation for a fixed branch set, at the cost of changing the potential. We therefore use the complete 16-branch formulation for conservative-force applications.} At the same time, the precision experiments show that throughput should not be reported independently of numerical deviation. FP32 and BF16 accelerate inference but introduce measurable energy differences relative to FP64. Training the reference checkpoint in FP64 provides a high-precision baseline from which these deployment-time tradeoffs can be quantified. \textcolor{black}{The crystal-relaxation benchmark extends the precision comparison beyond single-point inference. All 26 paired starting configurations converged in both FP32 and FP64, with closely agreeing final energies and cell geometries, including the compressed and expanded diamond-carbon systems.}

\textcolor{black}{Related adaptive-routing work includes the SCAN framework of Wang and You~\cite{wang2026scan}, which uses dynamic routing to model and interpret salt--solvent formulations and their conductivity. Our mechanism addresses a different task: it computes composition-conditioned weights over 16 fixed, dataset-specific energy heads of a shared atomistic GNN, rather than routing representations for electrolyte-property prediction. The branch-weighting experiments show that the MLP changes its preference among dataset-specific prediction heads according to elemental composition and atom count. Most observed preferences are consistent with the broad chemical scope of the selected training sources, although individual choices are not always intuitive. This initial implementation illustrates how the heads can be combined without manually assigning each material to a dataset; future work can test different composition descriptors and weighting networks to improve routing robustness.}\par

Several limitations remain to be addressed. Despite its broad chemical and structural coverage, the 16-source training corpus does not exhaust the diversity of atomic environments or electronic-structure approximations relevant to materials modeling. Composition-conditioned branch weighting is trained with source identity as the supervisory signal and therefore should not be interpreted as a calibrated uncertainty estimator. Finally, the first-principles cost comparison used for the billion-structure screening experiment is illustrative; actual DFT cost varies strongly with atom count, basis, convergence settings, and hardware. These limitations define concrete directions for subsequent work on fidelity calibration, uncertainty-aware routing, task-balanced sampling, and active selection of new first-principles data.

\section{Conclusions}
{
\textcolor{black}{
We presented an integrated exascale HydraGNN-based framework for training, adapting, and deploying multi-task atomistic GFMs on heterogeneous, multi-source, multi-fidelity data. MTL provides the architectural foundation for accommodating heterogeneous supervision, while scalable data management and distributed execution make it possible to train efficiently on hundreds of millions of atomistic structures across thousands of GPUs. Composition-conditioned weighting enables the deployment of MTL to predict properties of atomistic structures that do not necessarily belong to any individual training dataset, thereby promoting transferability across previously unseen chemistries.
Our experiments highlight the importance of considering computational efficiency alongside predictive accuracy throughout model development, training, and deployment. Large-scale HPO and fixed-budget continual training were instrumental in identifying PaiNN as a compact, effective architecture under realistic computational constraints. The downstream study similarly shows that transferability must be evaluated jointly with adaptation cost. HydraGNN achieves particularly strong results in the 150-structure Wiggle150 regime, competitive frozen transfer on QM9, and successful execution of all tested adaptation configurations across eight chemically diverse MS25 systems. On the latter benchmark, external full fine-tuning frequently exceeds the imposed two-hour budget, while specialized pretrained models achieve stronger predictive accuracy in selected tasks. These contrasting results establish that practical model selection depends on both predictive performance and time-to-solution, as well as pretraining coverage, target properties, and adaptation strategy.
Inference co-design further removes redundant work, provides up to a 33$\times$ throughput gain over the initial multi-head path, and enables 1.1 billion structures to be evaluated in 50 seconds on Frontier. 
Ultimately, HydraGNN provides an open workflow for developing scientific foundation models as complete model–data–systems artifacts, with predictive quality evaluated alongside adaptation cost, numerical precision, portability, and deployment throughput.
}
}

{
\textcolor{black}{
Future work will focus on cross-fidelity calibration, improving the robustness and interpretability of composition-conditioned aggregation, and developing more systematic criteria for selecting pretrained models and adaptation protocols. 
}
}

\section*{Supporting Information}
Supporting Information accompanies this article with detailed software environments, an expanded analysis of the runtime differences among continual-training finalists, and numerical-stability diagnostics for finite-precision behavior in the multi-branch HydraGNN model. \textcolor{black}{The verified PaiNN block dimensions, prediction-head structure, parameter counts, and optimizer-coverage distinction are documented in the Supporting Information; the principal configuration is summarized in Table~\ref{tab:lead_model}.}

\section*{Data Availability Statement}
The original atomistic datasets used for pretraining are publicly available from the repositories cited in Table~1 and the corresponding references. The pre-processed ADIOS2-formatted versions of all 16 training datasets, the trained HydraGNN model weights and checkpoint files, the lead PaiNN foundation model, and the parameters of the branch-weighting MLP are publicly available through the OLCF Data Constellation as part of the HydraGNN\_Predictive\_GFM\_2026 release~\cite{osti_2562660}.
HydraGNN is open source at \url{https://github.com/ORNL/HydraGNN}. The open-source code used for the HydraGNN--UMA--MACE zero-shot and fine-tuning comparisons reported in this article is available from the \texttt{uma-sota-comparison} branch at \url{https://github.com/allaffa/HydraGNN_GFM_FineTuning4Materials/tree/uma-sota-comparison}. Platform installation scripts used to construct the reported software
environments are maintained in the HydraGNN repository.

\section*{Funding}
This work was supported by the Artificial Intelligence Initiative through the Laboratory Directed Research and Development Program at Oak Ridge National Laboratory, managed by UT-Battelle, LLC, for the U.S. Department of Energy under contract DE-AC05-00OR22725, and by the U.S. Department of Energy Office of Science, Office of Advanced Scientific Computing Research and Office of Basic Energy Sciences through the SciDAC FORUM-AI project.

\section*{Acknowledgments}
Computing resources were provided by the Oak Ridge Leadership Computing Facility through ASCR Leadership Computing Challenge award LRN070; the Argonne Leadership Computing Facility through the Director's Discretionary award HydraGNN; and the National Energy Research Scientific Computing Center through award ``ScienceAtScale@NERSC'' ASCR-ERCAP0034735. These facilities are U.S. Department of Energy Office of Science user facilities.

\section*{Conflict of Interest}
The authors declare no competing interests.

\bibliographystyle{IEEEtran}
\bibliography{references}
\end{document}